\documentclass[a4paper,fleqn,usenatbib]{mnras}

\usepackage[T1]{fontenc}
\usepackage{ae,aecompl}

\usepackage{graphicx}	
\usepackage{amsmath}	
\usepackage{amssymb}	
\usepackage{txfonts}
\usepackage{color, colortbl}
\usepackage{enumitem}
\usepackage{yfonts}
\usepackage{caption}
\usepackage{nicefrac}
\usepackage{soul}

\newcommand{\be}{\begin{equation}}
\newcommand{\ee}{\end{equation}}

\def\tauvis{\tau_{\rm vis}}

\defcitealias{Friedmann2018}{FM18}
\def\FM18{\citetalias{Friedmann2018}}
\defcitealias{Williams2020}{W20}
\def\W20{\citetalias{Williams2020}}

\newcommand{\nodata}{\centering\arraybackslash}

\title[SN Ia delay time in clusters]{The delay time distribution of Type-Ia supernovae in galaxy clusters: the impact of extended star-formation histories}

\author[J. Freundlich and D. Maoz]{Jonathan Freundlich,$^{1,2}$\thanks{E-mails: jonathan.freundlich@astro.unistra.fr; maoz@astro.tau.ac.il}
Dan Maoz,$^{1 \star}$
\\
$^1$School of Physics and Astronomy, Tel Aviv University, Tel Aviv 69978, Israel\\
$^2$Observatoire Astronomique, Universit\'e de Strasbourg, CNRS, 11 rue de l'Universit\'e, 67000 Strasbourg, France
}

\date{Accepted 2021 February 4. Received 2021 January 25; in original form 2020 December 1.}

\pubyear{2020}

\begin{document}
\label{firstpage}
\pagerange{\pageref{firstpage}--\pageref{lastpage}}
\maketitle

\begin{abstract}
	
The delay time distribution (DTD) of Type-Ia supernovae (SNe~Ia) is important for understanding chemical evolution, SN~Ia progenitors, and SN~Ia physics. Past estimates of the DTD in galaxy clusters have been deduced from SN~Ia rates measured in cluster samples observed at various redshifts, corresponding to different time intervals after a presumed initial brief burst of star formation. A recent analysis of a cluster sample at $z=1.13-1.75$ confirmed indications from previous studies of lower-redshift clusters, that the DTD has a power-law form, ${\rm DTD}(t)=R_1 (t/{\rm Gyr})^\alpha$, with amplitude $R_1$, at delay $t=1~\rm Gyr$, several times higher than measured in field-galaxy environments. This implied that SNe~Ia are somehow produced in larger numbers by the stellar populations in clusters. This conclusion, however, could have been affected by the implicit assumption that the stars were formed in a single brief starburst at high $z$. Here, we re-derive the DTD from the cluster SN~Ia data, but relax the single-burst assumption. Instead, we allow for a range of star-formation histories and dust extinctions for each cluster. Via MCMC modeling, we simultaneously fit, using stellar population synthesis models and DTD models, the integrated galaxy-light photometry in several bands, and the SN~Ia numbers discovered in each cluster. With these more-realistic assumptions, we find a best-fit DTD with power-law index $\alpha=-1.09_{-0.12}^{+0.15}$, and amplitude $R_1=0.41_{-0.10}^{+0.12}\times 10^{-12}~{\rm yr}^{-1}{\rm M}_\odot^{-1}$. We confirm a cluster-environment DTD with a larger amplitude than the field-galaxy DTD, by a factor $\sim2-3$ (at $3.8\sigma$). Cluster and field DTDs have consistent slopes of $\alpha\approx-1.1$.

\end{abstract}




\section{Introduction}
\label{section:introduction}

Type-Ia supernovae (SNe Ia) are explosions involving the thermonuclear combustion of white-dwarf stars, although the precise identity of the exploding systems, the processes leading to explosion, and the development of the explosion itself, are all far from clear yet \citep[see][for reviews]{Maoz2014, LivioMazzali2018}. Together with their more-numerous cousins, core-collapse supernovae (CC-SNe), SNe Ia are the production factories for the elements from oxygen to the iron-peak elements, they are also sources of kinetic-energy feedback that plays a role in galaxy formation, and their remnants are the acceleration sites of cosmic rays. Among the observable physical properties of the SN Ia population, the delay-time distribution (DTD) of SNe Ia has been useful for investigating both the SN Ia progenitor question and the contribution of SNe Ia to the chemical and structural evolution of galaxies. As its name implies, the DTD is the distribution of times elapsed between the formation,  at time zero, of a hypothetical stellar population, of unit mass, and the explosion, as SNe Ia, of some of its white-dwarf remnants \citep[see][for a review on the DTD and its measurement]{Maoz2012}. The DTD, usually expressed as a SN Ia rate per formed stellar mass as a function of delay time, embodies the "impulse response" of the SN Ia phenomenon, and can provide clues relevant for studying all of the issues mentioned above.

The DTD is directly linked to the lifetimes of SN Ia progenitor systems, and it could, in principle, vary with environment, cosmic time, metallicity, and initial mass function (IMF). The past decade has seen numerous efforts to measure the DTD, using diverse observational methodologies and analyses, applied to SNe Ia in different environments and times (field galaxies, cluster galaxies, low and high redshift).
For field galaxies, the DTD has been estimated by deconvolving the volumetric SN~Ia rate, as a function of redshift, from the cosmic star formation history (SFH), where the DTD is the convolution kernel. Similarly, the field-galaxy DTD has been also derived from comparison of SN Ia rates in galaxies to their individual galaxy SFHs.
Many of these studies have converged toward a DTD with a power-law time dependence, $t^\alpha$, with $\alpha\sim -1$, between delays $t\sim 40$~Myr and a Hubble time. The DTD is assumed to be zero at $0\leq t \lesssim 40$~Myr, since a stellar population has not yet formed any white dwarfs, essential for a SN Ia explosion, during that interval. Observations have yet to constrain well the DTD at short delays, of order $\sim 100$~Myr and below. The latest analyses of SN Ia rates in field galaxies \citep[e.g.,][]{Maoz2017} find a DTD with a power-law index $\alpha=-1.1\pm 0.1$, and a time-integrated (from 40~Myr to 13.7~Gyr) number of SNe Ia per formed stellar mass\footnote{As in most recent works on the subject, for the calculation of formed stellar mass, based on observed light from a stellar population, we will assume in this paper a \cite{Kroupa2001} IMF. We note that this assumption is, more than anything, a calibration convention that facilitates comparisons of diverse measurements of SN Ia rates, that are all actually normalised relative to the observed stellar luminosity of their host-galaxy populations, rather than to their current or formed mass, which is not directly observed.}, $N_{\rm Ia}/M_*=(1.3\pm 0.1)\times 10^{-3}$~M$_\odot^{-1}$. 

In galaxy clusters, whose stars are assumed to have been formed in single brief bursts at  redshifts $z_f\sim3-4$ (more on this assumption further below), the DTD has been recovered simply by measuring the SN~Ia rate in galaxy-cluster samples as a function of their redshift, with different redshifts corresponding to different delays since the epoch of the  single star-formation burst.
Most recently, \citet[][\citetalias{Friedmann2018}]{Friedmann2018} searched for SNe Ia in a sample of clusters at $z=1.13-1.75$ monitored by the {\it Hubble Space Telescope} (HST). Fitting a power-law DTD to the the SN Ia rate in this high-$z$ sample jointly with the rates measured by previous studies of lower-redshift samples, FM18 found best-fit parameters $\alpha=-1.30^{+0.23}_{-0.16}$ (i.e. a power-law index somewhat, although not significantly, steeper than seen in field environments) and a Hubble-time-integrated SN Ia production efficiency, $N_{\rm Ia}/M_*$, at least several times larger than the value of $N_{\rm Ia}/M_*$ measured by field-SN Ia studies. Both of these results confirmed indications deduced from previous, lower-$z$, cluster samples \citep[e.g.,][]{Maoz2010, Barbary2012}.

The high normalisation of the DTD of SNe Ia in galaxy clusters is particularly intriguing. It implies that something in the mode of star-formation or stellar evolution that took place in these environments is conducive to an overall larger production of SNe Ia, at all delays. FM18 speculated that, at the root of the phenomenon, could be, e.g., the formation of a higher fraction of close binaries in the mass range of stars that are the progenitors of SNe Ia; or perhaps a "middle-heavy" IMF, with an excess of intermediate-mass stars, that evolve into white dwarfs in the mass range that explode as SNe Ia. There is already mounting evidence for IMF pecularities in massive elliptical galaxies, the very same type that dominate galaxy clusters, even if there is yet disagreement about the driving agents of these peculiarities (galaxy mass, metallicity, age, etc.). 
\citet{vanDokkum2010}, \citet{Treu2010}, \citet{Conroy2012}, \citet{Conroy2017}, \citet{Cappellari2012,Cappellari2013}, \citet{Lyubenova2016}, \citet{Davis2017}
and others have deduced the existence of a "bottom-heavy" IMF in such galaxies: the IMF slope, rather than becoming shallower at stellar masses below $\sim 0.5$~M$_\odot$, continues with the "Salpeter" slope down to the hydrogen-burning limit at  $\sim 0.1$~M$_\odot$. The stellar atmospheres in such galaxies also have a high ratio of $\alpha$ elements to iron [$\alpha$/Fe], compared to the stars in lower-mass and disk galaxies, and compared to the disk stars in the Milky Way \citep[e.g.,][]{Conroy2014}. Since the bulk of the $\alpha$ elements are produced by CC-SNe from massive stars, while iron is synthesized both in SNe Ia and CC-SNe \citep[with roughly equal contributions to the universal iron budget, see][]{Maoz2017}, 
a peculiar [$\alpha$/Fe] ratio is therefore suggestive of IMF variations at the high end of the IMF: a change in the high-mass IMF slope might change the resulting mix of different CC-SN types, thus changing the integrated [$\alpha$/Fe] from CC-SNe; and/or a change in the ratio of high-mass stars (that explode as CC-SNe) and intermediate-mass stars (some of whose white-dwarf descendants produce SNe Ia), could also affect [$\alpha$/Fe]. \citet{Hallakoun2020} have recently discovered a Salpeter-like bottom-heavy IMF in yet another ancient, high [$\alpha$/Fe], environment---the "metal-poor" or "blue" halo of the Milky Way, thought to be the stellar debris of a galaxy accreted by our Galaxy 10 Gyr ago. 
Although the low-mass end of the IMF is unlikely to be relevant for SNe Ia, whose white-dwarf progenitors derive from intermediate-mass main-sequence stars, these latest results show that IMF variations are possible among different galaxy populations or environments. The high-normalised DTD measured in galaxy clusters could thus be another indicator of an IMF that is peculiar to star formation in early-type galaxies, cluster environments, or early cosmic epochs. It could also be related to the long-known excess of the iron-to-stellar mass ratio in clusters, compared to expectations from SN yields \citep[e.g. see][for a recent analysis]{Ghizzardi2020}. 

However, the measurement of the DTD in galaxy clusters, and the results described above, have hinged on the assumption that the stars in galaxy clusters were all formed in a short burst at high redshift, with negligible ensuing star forming thereafter. This assumption has permitted to associate directly the SN Ia rate per unit mass, measured in a sample of clusters at redshift $z$, with the DTD at a delay $t_{\rm delay}$ corresponding to the time elapsed between the formation redshift, $z_f$, and the cluster redshift $z$. The short-burst assumption is likely justified, at least  when analyzing clusters at $z\lesssim 1$. Spectral analyses of the stellar populations of cluster galaxies have long shown that the bulk of their stars were formed quickly at $z_f\sim 3-4$ \citep[e.g., ][]{Daddi2000,Stanford2005,Eisenhardt2008,Snyder2012,Stalder2013,Andreon2014, Andreon2016}. 
Most recently, \citet{Salvador-Rusinol2020} have analysed the ultraviolet-to-optical absorption lines in the high-signal-to-noise stacked spectra of tens of thousands of massive early-type galaxies at $z\sim 0.4$, and have set a stringent limit of $<0.5$\% on the fraction of these galaxies' stellar mass formed in the prior 2 Gyr. 

A concern, however, for cluster galaxies observed at $z\sim 1-2$, corresponding to intervals of just $\sim 1-4$~Gyr after initial star formation, is that for such short delays the single-short-burst assumption might be unjustified, and could therefore result in an incorrectly determined DTD. Significant ongoing star formation in the cluster galaxies at the epoch of observation could, on the one hand, cause the DTD delay time effectively being probed by the measured SN Ia rate to be actually smaller than the delay implicit in the single short burst assumption, biasing high the DTD (since a shorter delay, in a monotonically decreasing DTD, would actually be probed). On the other hand, neglecting the presence of a young population, with a low mass-to-light-ratio, in the cluster galaxies would lead to an overestimate of the formed stellar mass and to an underestimate of the SN Ia rate, and hence would bias the DTD low. To understand which, if either, of these biases are at work, and to obtain a more robust estimate of the DTD form and amplitude in clusters, the single-burst assumption must be relaxed.

In this paper, we re-analyse the SN Ia data in \citetalias{Friedmann2018} and the cluster SN Ia rates measured by previous studies, to find the observational constraints on the DTD in the presence of {\it extended} (rather than single-burst) star formation histories, but only those histories that are consistent with the observed luminosities and spectral energy distributions of the monitored clusters. To this end, we re-measure and revise the multi-band near-infrared (rest-frame optical) photometry of the galaxy clusters in \FM18. When modeling the data, we compare the predicted numbers of SNe Ia in every cluster in \FM18 to the actual number of SNe Ia discovered in that cluster (rather than to the number of SNe Ia in the sample as a whole, as in \citetalias{Friedmann2018}), thus utilising more information. The SN Ia numbers are also slightly revised, in view of recent followup observations of the \FM18 sample that have led to some clarifications of SN classification and cluster membership of host galaxies. In addition we revise some of the estimates of formed stellar mass, and hence of the mass-normalised rates, in lower-redshift cluster SN samples, based on improved modeling of their reported stellar luminosities.   

\section{Revised cluster stellar fluxes, supernova numbers, and supernova rates}
\label{section:fluxes}

\subsection{Cluster stellar fluxes}
\label{subsection:photometry}

\FM18 measured and reported the net cluster-galaxy fluxes in three near-infrared broad photometric HST bands: F105W (centred at wavelength 
$\sim 1.05~\mu$m), F140W (1.40~$\mu$m),  and F160W (1.60 $\mu$m). At the typical cluster redshifts, these bands correspond roughly to the rest-frame optical $B$, $V$, and $R$ bands. In their analysis, \FM18 used only the flux they measured in the reddest band flux, F160W, to estimate, under the early-short-burst assumption, the formed stellar mass of each cluster, by comparing the observed flux to predictions of spectral synthesis models of a single burst at a redshift $z_f=3$ or 4. Without spectroscopy of all galaxies in the field of view, it is impossible to clearly discriminate cluster galaxies from foreground and background galaxies. Following previous studies, \FM18 therefore performed a statistical subtraction of the foregrounds and the backgrounds. For this purpose, \FM18 used HST images of the Hubble Ultra-Deep Field \citep[HUDF;][]{Beckwith2006}, obtained with the same instrumental setup, in the role of the typical extragalactic sky in a non-galaxy-cluster direction. However, in the course of the present work, we realized that the measurements of \FM18 for some of the clusters in some of the bands are erroneous, apparently due to over-subtraction of foreground galaxy light. We have therefore re-measured the photometry of all of the \FM18 clusters in their available bands.

For our new photometry, we first measured the total counts in a cluster image within a circular aperture, centred on a cluster's brightest cluster galaxies (BCGs), with an aperture radius of $51$ arcseconds, but up to $55$ arcseconds for some of the clusters. The aperture radius for each cluster was chosen to avoid including within the aperture, if possible, bright foreground stars and galaxies near the aperture borders, whose subsequent subtraction would add noise to the photometric measurement. The aperture covers roughly 50\% of the WFC3-IR detector's image area, comparable to the "fully time-covered area" in each cluster field that was included in all HST epochs and all bands of a cluster, due to the image plane rotation from epoch to epoch in the course of the 2-year SN Ia survey observations (see \FM18). Only SNe Ia discovered within this area were included by \FM18 in the SN Ia census counted toward the SN Ia rate, and therefore the light only from galaxies within this area needs to be integrated for estimating a cluster's stellar luminosity and stellar mass that normalise the SN Ia rate. Next, we identified, within the aperture, foreground Galactic stars (easily identifiable in HST images via their diffraction spikes) and bright foreground galaxies that are brighter than a cluster's BCGs. We photometered these stars and galaxies using small apertures ($\sim 1$~arcsecond, or as appropriate), subtracting the background counts estimated in annuli around the small apertures, and subtracted the foreground-star and bright-galaxy counts from the total counts of the large cluster aperture. Finally, we estimated the remaining non-cluster backgrounds and foregrounds by placing $13$-arcsecond-radius apertures at diverse, relatively sparsely populated, locations around the image edges, avoiding stars and obvious galaxy concentrations (either cluster- or foreground galaxies). These remaining backgrounds/foregrounds result from: residual dark current and residual cosmic-ray events in the detector; airglow; zodiacal light; faint, both individually detected and undetected (below the flux detection limit) field (i.e. non-cluster) galaxies, and integrated extragalactic background light. The counts per pixel in these marginal-region apertures were used to estimate and subtract the total foreground and background counts within the large cluster aperture. The variations in counts from one edge region to another were used to estimate the uncertainty in the foreground/background subtraction, which dominates the uncertainty in the final cluster-light photometry. This method of estimating a cluster's total stellar flux, by adding all counts within the cluster aperture, automatically includes all of the individually undetected faint cluster galaxies (i.e. galaxies with flux below the detection limit) plus any intra-cluster starlight, if it exists. Our new measurement procedure is essentially very similar to one of the two photometry approaches used by \FM18, except for our field-edge-based foreground/background counts estimate, which replaces the HUDF-based estimate in \FM18 and which, as noted, sometimes resulted in over-subtraction, and hence an underestimate of the cluster stellar fluxes. Table~\ref{table:fluxes} lists our new photometric measurements, which supersede those published in table 5 of \FM18. 

\begin{table*}
	\caption{Revised cluster galaxy fluxes and hosted SN Ia numbers.}
	\begin{tabular*}{\linewidth}{l@{\extracolsep{\fill}}l@{\extracolsep{\fill}}l@{\extracolsep{\fill}}l@{\extracolsep{\fill}}l@{\extracolsep{\fill}}l@{\extracolsep{\fill}}l@{\extracolsep{\fill}}l@{\extracolsep{\fill}}l@{\extracolsep{\fill}}l@{\extracolsep{\fill}}l}
		\hline
		\hline
		\noalign{\vskip 0.5mm} 
        Cluster name & $z$ & $f_{\rm F105W}$ &  $f_{\rm F140W}$ &  $f_{\rm F160W}$ & $N_{\rm Ia}$ & $N_{\rm Ia?}$ & $\tauvis$ & $\eta_{-40}$ & $\eta_0$ & $\eta_{+20}$\\
        (1) & (2) & (3) & (4) & (5) & (6) & (7) & (8) & (9) & (10) & (11) \\
		\noalign{\vskip 0.5mm} 
		\hline
		\noalign{\vskip 0.5mm} 
		IDCS1426 & 1.75 & $0.50 \pm 0.45$ & $0.68 \pm 0.10$ &  $0.53 \pm 0.11$ & 1 & 1 & 52, 89 & 0.71, 0.78 & 0.86, 0.87 & 0.75, 0.79 \\
		%
		ISCS1432 & 1.40 & $0.70 \pm 0.18$ & $0.97 \pm 0.06$ & $0.81 \pm 0.11$ & 1 & 0 &104, 87 & 0.88, 0.88 & 0.93, 0.96 & 0.86, 0.86\\
		%
		MOO1014 & 1.23$^\dagger$ & $1.56 \pm 1.20$ & $3.02 \pm 0.19$ & $1.38 \pm 0.17$ & 2 & 0 & 98, 88 & 0.93, 0.92 & 0.99, 0.98 & 0.89, 0.89 \\
		%
		MOO1142 & 1.19 & $1.94 \pm 0.70$ & $1.87 \pm 0.13$ & -- & 0 & 0 & 89 & 0.87 & 0.93 & 0.88 \\
		SPARCS0224 & 1.63 & $1.00 \pm 0.30$ & $0.89 \pm 0.16 $ & $0.67 \pm 0.26$ & 1 & 1 &67 & 0.72 & 0.83 & 0.81 \\
		SPARCS0330 & 1.63 & $0.65 \pm 0.65$ & $0.76 \pm 0.38$ & $0.55 \pm 0.15$ & 0 & 1 & 112 & 0.77 & 0.87 & 0.85 \\
		SPARCS1049 & 1.70 & $1.30 \pm 0.25$ & -- & $0.80 \pm 0.14 $ & 0  & 0 & 41 & 0.56 & 0.63 & 0.62 \\
		SPARCS0035 & 1.34 & $1.28 \pm 0.25$ & $1.22 \pm 0.18$ & -- & 0 & 1 & 75, 45 & 0.89, 0.88 & 0.95, 0.97 & 0.87, 0.88 \\
		%
		SPT0205 & 1.58 & $0.90 \pm 0.25$ & $1.14 \pm 0.13$ & $1.09 \pm 0.11$ & 4 & 1 & 310 & 0.93 & 0.99 & 0.88 \\
		SPT2040 & 1.48 & $1.43 \pm 0.35$ & $2.13 \pm 0.23$ & $1.11 \pm 0.30$ & 3 & 0 & 101, 42 & 0.85, 0.85 & 0.96, 0.94 & 0.91, 0.84 \\
		%
		SPT2106 & 1.13 & $3.55 \pm 0.30$ & $1.98 \pm 0.13$ & -- & 1 & 0 & 90,71 & 0.93,0.95 & 0.98,0.98 & 0.90,0.91 \\
		%
		XMM44 & 1.58 & $1.30 \pm 0.10$ & $1.32 \pm 0.08$ & $1.15 \pm 0.13$    & 1 & 0 & 70 & 0.69 & 0.79 & 0.77 \\
		\hline
	\end{tabular*}
	\begin{minipage}{1\linewidth}
	\textbf{Notes.} 
	(1) Cluster ID; 
	(2) Cluster redshift; 
	(3) -- (5) Total measured galaxy cluster flux within the full-time-coverage area of the field in the F105W, F140W, and F160W bands, when available, in $10^{-16}~\rm erg~cm^{-2}~s^{-1}~\AA^{-1}$;
	(6) (7) Number of likely and possible cluster SNe Ia, respectively;
	(8) Cluster rest-frame monitoring time for each season, in days; 
	(9) -- (11) Detection-completeness factors for the periods before, during, and after each observing season, respectively, from \FM18. When more than one season to a cluster, the two season's monitoring times and completeness factors are listed. %
	$^\dagger$Revised from $z=1.27$, based on spectroscopy by \protect\citet{Williams2020}.
    \end{minipage}
	\label{table:fluxes}
\end{table*}

\subsection{Cluster SN Ia numbers}
\label{subsection:NSNIa}

In their analysis of the HST SN survey data, \FM18 discovered 29 transient events. Each transient's multi-band light curves were  
fit with SN Ia model curves, to gauge whether or not the transient was a SN Ia in the cluster. At the end of this process, \FM18 reported that 11 events were likely SNe Ia in cluster galaxies, within the fully time-covered areas of the cluster fields, and a further 4 events were deemed possible cluster SNe Ia. At the time of this writing, the Supernova Cosmology Project (SCP) team, who proposed and led the HST observations carried out in 2014-2016, have not yet published the results of their own search for SNe in these data. However, \citet[][\W20]{Williams2020} have reported the results of ground-based spectroscopic observations by the SCP of some of the host galaxies of the transients that the group discovered in the southern-hemisphere clusters in the sample. We have attempted to cross-match all of the transients mentioned in \W20 with all of the transients previously reported by \FM18. That attempt is summarized in Table~\ref{table:cross-match}.
Some of the transients found by \FM18 are not mentioned in \W20 (even in those clusters for which \W20 show images with the SCP-detected transients indicated), and vice versa. The non-mutual detections may have resulted from differing estimates, between the two groups, regarding the significance and reality of those detections. For example, some of the transients identified by \W20 but not by \FM18 are located near or on the nuclei of bright galaxies, some of them nuclei spectroscopically identified as active galactic nuclei (AGN), which are often variable. Nonetheless, the majority of the cluster SN Ia candidates are detected and similarly classified, based on their light curves, by the two studies. The spectroscopy of the host galaxies (and in one case, of the SN itself) by \W20 can confirm or refute the cluster membership of some of the SN Ia candidates. We use the \W20 results to slightly revise, for three clusters, the estimated \FM18 numbers of SNe Ia that the clusters hosted. The revised SN Ia numbers are listed in Table~\ref{table:fluxes}, and supersede those in table 5 of \FM18. The cases of clusters with re-assessed SN Ia numbers follow below.

In the cluster MOO1014, the galaxy spectroscopy by \W20 leads them to revise the cluster redshift, from $z=1.27$ down to $z=1.23$. The event SNFM10, reported by \FM18 and well-fit as a SN Ia in the cluster, is not mentioned in \W20. Transient SNFM11, deemed by \FM18 a cluster SN Ia, corresponds in \W20 to SCP15C02. \W20 find that its host is a foreground galaxy at $z=0.7506$, and therefore we no longer count it as a cluster SN Ia. SNFM13 (=SCP14C01 in \W20) was a hostless event marginally fit as a SN Ia, which \FM18 classified as most probably a foreground event. However, \W20 present a "live" spectrum of the event itself and they conclude that it is likely a SN Ia at the cluster redshift. SCP15C04 is an event not  reported by \FM18, near the nucleus of its host galaxy, which may have affected the evaluation of its significance. \W20 confirm the galaxy as a cluster member and list the event as a possible SN Ia. Given its non-detection by \FM18 at its problematic location, and its uncertain classification by \W20, we do not adopt this event into our cluster SN Ia sample. The remaining transients in cluster MOO104, whether reported by \FM18, by \W20, or by both, are all non-cluster on non-SNIa events (see Table~\ref{table:cross-match}). In the balance, one of this cluster's assumed likely SNe Ia in the \FM18 sample drops out (SNFM11), and one enters it as such (SNFM13). Together with SNFM10, the number of likely SNe Ia in this cluster remains unchanged at 2.

In SPARCS0224, \FM18 discovered transient SNFM15 in a spectroscopically confirmed cluster host galaxy, which they classified as a possible SN Ia. \W20 do not report this event. An additional transient, SNFM16 (=SCP16I02), was considered by \FM18 a likely cluster SN Ia. \W20 measure for its host a tentative redshift of $z=1.62$, consistent with cluster membership, but they state that this redshift is not secure. For this cluster as well, the cluster SNe Ia numbers remain unchanged, at one possible and one likely SN Ia. 

In cluster SPT0205, \FM18 discovered two transients, SNFM22 (=SCP15A04) and SNFM23 (=SCP16A04) in the same host galaxy, separated by about a year in the observer frame. The light curves of both were well-fit by \FM18 with SN Ia model light curves at the cluster redshift. A spectrum by \W20 confirms the galaxy is in the cluster. Similarly, the host of SNFM24 (=SCP15A05), classified by \FM18 as a cluster SN Ia, is confirmed by \W20 as a cluster galaxy. For SNFM25 (=SCP15A01), also counted as a cluster SN Ia by \FM18, the host galaxy was too faint for \W20 to secure a redshift. In addition, and not detected by \FM18, \W20 report SCP15A06 as a likely SN Ia, with position nearly coincident with the nucleus of its host, which has a redshift $z=1.3128$. This redshift is consistent with, but somewhat lower than, the redshifts of the other host galaxies in this cluster measured by \W20, $z=1.322$ to $z=1.340$, raising the possibility that this galaxy could be foreground to the cluster. All things considered, we count SCP15A06 as a "possible" cluster SN Ia, in addition to the 4 likely SNe Ia already considered in \FM18 in this cluster.

In the cluster SPT2040, \FM18 discovered event SNFM26 (=SCP15E04 in \W20) in a faint host galaxy. \FM18 and \W20 both consider this a likely cluster SN Ia event. No spectrum is reported by \W20. Not detected by \FM18, however, are two transients, SCP15E06 and SCP15E07, reported by \W20 as likely SNe Ia, with their hosts confirmed as cluster members. We therefore add these two events to SNFM26, for a total of 3 likely SNe Ia in this cluster. We do not include in our tally an additional event in this cluster, SCP15E03, mentioned by \W20 as a possible SN Ia, and coincident with the nucleus of a galaxy lacking a measured redshift.

For the cluster SPT2106, \W20 report that SCP16D01 (not detected by \FM18), deemed a likely SN Ia, is hosted by a galaxy at the cluster redshift. We therefore raise the number of likely SNe Ia in this cluster from zero to one.
Finally, in cluster XMM44, \FM18 classified SNFM29 (=SCP15G01 in \W20) as a likely cluster SN Ia. The host spectrum by \W20 confirms its cluster membership.

For the full \FM18 cluster sample, the total number of cluster SNe Ia is thus revised from 11 to 14 likely SNe Ia, and from 4 to 5 possible SNe Ia. 
\begin{table*}
	\caption{Cross-match between transients in \FM18 and \W20.}
	\begin{tabular*}{\linewidth}{l@{\extracolsep{\fill}}l@{\extracolsep{\fill}}l@{\extracolsep{\fill}}l}
		\hline
		\hline
		\noalign{\vskip 0.5mm} 
        Cluster & Name in \FM18 & Name in \W20 & Notes\\
		\noalign{\vskip 0.5mm} 
		\hline
		\noalign{\vskip 0.5mm} 
IDCS1426  &  SNFM01   &   \nodata  &   possible foreground SN Ia  \\
          &  SNFM02   &   \nodata  &   hostless, near field edge \\
          &  SNFM03   &   \nodata  &   possible cluster SN Ia \\
          &  SNFM04   &   \nodata  &   non-SN Ia \\
          &  SNFM05   &   \nodata  &   foreground non-SN Ia \\
          &  SNFM06   &   \nodata  &   likely cluster SN Ia \\
\hline
\noalign{\vskip 0.5mm} 
ISCS1432  &  SNFM07   &   \nodata  &   likely cluster SN Ia \\
          &  SNFM08   &   \nodata  &   hostless, near field edge  \\
          &  SNFM09   &   \nodata  &   not discovered in F140W search band \\
\hline
\noalign{\vskip 0.5mm} 
MOO1014   &  SNFM10   &   \nodata  &   likely cluster SN Ia \\
          &  SNFM11   &   SCP15C02 &   foreground CC-SN at $z=0.7506$ \\
          &  SNFM12   &   SCP16C03 &   lensed background SN Ia at $z=2.22$ \\
          &  SNFM13   &   SCP14C01 &   likely cluster SN Ia, based on "live" \W20 spectrum \\
          &  \nodata  &   SCP15C01 &   SN Ia in foreground at $z=0.9718$ \\
          &  \nodata  &   SCP15C03 &   non-SN Ia, possibly $z=1.01$ \\
          &  \nodata  &   SCP15C04 &   cluster host; \W20: "possible SN Ia" \\
          &  \nodata  &   SCP16C01 &   foreground non-SN Ia, $z=9739$ \\
          &  \nodata  &   SCP16C02 &   non-SN Ia, no host redshift \\
\hline
\noalign{\vskip 0.5mm} 
MOO1142   &  SNFM14   &   \nodata  &   hostless, non-SN Ia, peculiar light curve \\
\hline
\noalign{\vskip 0.5mm} 
SPARCS0224 & SNFM15   &   SCP15I01  &   possible SN Ia in confirmed cluster member \\
           & SNFM16   &   SCP15I02  &  likely SN Ia, host at cluster redshift but not secure \\
\hline
\noalign{\vskip 0.5mm} 
SPARCS0330 & SNFM17   &   SCP16H01 &   possible cluster SN Ia \\
           & SNFM18   &   \nodata  &   non-SN Ia  \\
\hline
\noalign{\vskip 0.5mm} 
SPARCS0035 & SNFM19   &   \nodata  &   hostless, non-SN Ia \\
           & SNFM20   &   \nodata  &   possible cluster SN Ia \\
\hline
\noalign{\vskip 0.5mm} 
SPT0205    & SNFM21   &   SCP16A03 &   non-SN Ia \\
           & SNFM22   &   SCP15A04 &   likely SN Ia in confirmed cluster galaxy \\
           & SNFM23 &     SCP16A04 &   likely SN Ia in confirmed cluster galaxy \\
           & SNFM24  &    SCP15A05 &   likely SN Ia in confirmed cluster galaxy \\
           & SNFM25   &   SCP15A01 &   likely cluster SN Ia \\
           & \nodata  &   SCP15A02 &   non-SN Ia, $z=0.8966$ \\
           & \nodata  &   SCP15A03 &   cluster SN Ia, but near field edge \\
           & \nodata  &   SCP15A06 &   possible SN Ia, $z=1.3128$, possibly cluster foreground \\
           & \nodata  &   SCP15A07 &   non-SN Ia at $z=0.5015$ \\
           & \nodata  &   SCP16A01 &   non-SN Ia, $z=0.4998$ \\
           & \nodata  &   SCP16A02 &   non-SN Ia, $z=0.7971$ \\
\hline
\noalign{\vskip 0.5mm} 
SPT2040    & SNFM26   &   SCP15E04 &   likely cluster SN Ia \\
           & SNFM27   &   SCP15E02 &   non-SN Ia \\
           & \nodata  &   SCP15E01 &   non-SN Ia, $z=0.8398$ \\
           & \nodata  &   SCP15E03 &   possible SN Ia, on nucleus of galaxy with no redshift \\
           & \nodata  &   SCP15E05 &   non-SN Ia, on nucleus of bright galaxy \\
           & \nodata  &   SCP15E06 &   likely SN Ia in confirmed cluster galaxy \\
           & \nodata    & SCP15E07 &   likely SN Ia in confirmed cluster galaxy \\
           & \nodata   &  SCP15E08 &   possible SN Ia in broad-lined AGN at $z=2.02$ \\
           & \nodata   &  SCP16E02 &   non-SN Ia, $z=0.9435$ \\
\hline
\noalign{\vskip 0.5mm} 
SPT2106    & SNFM28    &  \nodata &    non-SN Ia, near field edge \\
           & \nodata   &  SCP15D01 &   foreground SN Ia, $z=0.5682$, near nucleus of bright galaxy \\
           & \nodata   &  SCP15D02 &   possible SN Ia on the nucleus of a cluster-member AGN  \\
           & \nodata   &  SCP15D03 &   possible SN Ia near nucleus of possibly foreground galaxy at $z=1.1130$ \\
           & \nodata   &  SCP15D04 &   possible SN Ia, field edge, no clear host \\
           & \nodata   &  SCP16D01 &   likely SN Ia in confirmed cluster galaxy \\
           & \nodata   &  SCP16D02  &  non-SN Ia, near field edge \\
           & \nodata   &  SCP16D03 &   non-SN Ia, $z=0.6118$ \\
\hline
\noalign{\vskip 0.5mm} 
XMM44      & SNFM29    &  SCP15G01 &   likely SN Ia in confirmed cluster galaxy \\		
		\hline
	\end{tabular*}
	\label{table:cross-match}
\end{table*}


\subsection{Revised SN Ia rates}
\label{subsection:revised_SNIa}

The observed SN Ia rate per formed stellar mass was calculated in \citetalias{Friedmann2018} as 
\be
\label{eq:Robs}
R_{\rm Ia,M} = \frac{N_{\rm Ia, tot}}{\sum\limits_{i} M_{i} \eta_{i} \tau_{{\rm vis},i}}, 
\ee
where $N_{\rm Ia, tot}$ is the total number of transients in a cluster sample that were classified as SNe Ia, $M_i$ is the  formed stellar mass under the assumption of a single instantaneous burst, $\tau_{{\rm vis},i}$ are the rest-frame monitoring times, and $\eta_i$ are the detection-completeness factors. The sum in the denominator is simultaneously over the different clusters, their observing seasons, and the periods before, during, and after each season. As our revised numbers for the \citetalias{Friedmann2018} sample now include 14 likely and 5 possible SN Ia, we adopt $N_{\rm Ia, tot} = 16.5 \pm 2.5 {\rm 
~(systematic~error)~} \pm 4.1 {\rm 
~(Poisson~error)}$.

To re-estimate the SNe Ia rate and error, as implied by the \FM18 sample, we calculate $R_{\rm Ia,M}$ with 10,000 realisations. In each realisation, we draw a value of $N_{\rm Ia, tot}$ from a distribution centred on it mean value, and having width according to its systematic and Poisson uncertainties.

In each realisation and for each cluster, we also draw a delay time, uniformly spread within the range corresponding to formation redshifts between $z_f=3-4$. 
The formed stellar mass is determined, as in \FM18, i.e., by comparing the total stellar luminosity of a cluster in its reddest available band, and its uncertainty, to the luminosity in that band predicted by spectral-population synthesis (SPS) modeling, for a single burst given the delay time (the delay time range constituting a further source of uncertainty for the derived mass). 

For the SPS generation we use the \texttt{P\'egase.3} SPS code \citep{Fioc1997, Fioc1999, Fioc2019} instead of \texttt{Starburst99} \citep{Leitherer1999,Leitherer2010, Leitherer2014, Vazquez2005} used by FM18. We assume a constant Solar stellar metallicity $Z=0.020$, a \cite{Kroupa2001} IMF, and Padova+AGB stellar evolution tracks \citep[cf.][section 2.2.2 and references therein]{Fioc2019}. We have verified that we obtain essentially identical model spectra with  \texttt{P\'egase.3}  and with \texttt{Starburst99}, for instantaneous bursts of star formation with the same input parameters.

For the full \FM18 sample, we obtain a SNe Ia rate per unit mass of $R_{\rm Ia,M}=24.6\pm 7.1 \times 10^{-14}~\rm yr^{-1}M_\odot^{-1}$. %
We also re-derive the SN Ia rate per unit rest-frame B-band stellar luminosity, $R_{\rm Ia,L_B}$, which is calculated the same as $R_{\rm Ia,M}$, except that the formed stellar mass, $M_i$, is replaced by the $B$-band luminosity of each cluster, $L_{B,i}$, in Eq.~(\ref{eq:Robs}). The $B$-band luminosity is deduced from the \texttt{P\'egase.3} spectrum. We find $R_{\rm Ia,L_B}=0.64\pm 0.19 \times 10^{-12}
~\rm yr^{-1}L_\odot^{-1}$, as listed in Table~\ref{table:rates}. 
The delay time range corresponding to a formation redshift $z_f=3-4$ for the  mean redshift of the sample is $2.2-2.8$ Gyr.

Splitting the \citetalias{Friedmann2018} sample into two equal subsamples, one composed of the six lower-$z$ clusters ($z=1.1-1.4$) and the other of the six higher-$z$ clusters ($z=1.4-1.75$), the number of SN Ia in the lower-$z$ sample is $N_{\rm Ia, low} = 9.0 \pm 1.0 {\rm 
~(systematic)~} \pm 3.0 {\rm 
~(Poisson)}$ and that in the higher-$z$ sample $N_{\rm Ia, high} = 7.5 \pm 1.5 {\rm 
~(systematic)~} \pm 2.7 {\rm 
~(Poisson)}$. 
The corresponding SN Ia rates are respectively  $19.5 \pm 6.8$ and $37.0 \pm 15.5 \times 10^{-14}~\rm yr^{-1}M_\odot^{-1}$ per formed stellar mass, and 
$0.57 \pm 0.20$
and 
$0.74\pm 0.31 \times 10^{-12}
~\rm yr^{-1}L_\odot^{-1}$  per unit $B$-band luminosity. 
The delay-time ranges probed by the two subsamples are respectively $2.7-3.2$ and $1.8-2.4$ Gyr. 
The revised full-sample measurements of the SN Ia rates and those for the two subsamples are all within $\lesssim 10\%$ ($<0.5\sigma$) of the \citetalias{Friedmann2018} values.

In the studies of cluster SN Ia rates previous to \citetalias{Friedmann2018}, as compiled by \cite{Maoz2017} and listed also in table 6 of \FM18, cluster stellar masses were derived indirectly from the observed cluster luminosities. The color of a cluster stellar population was either measured from multi-band data, or assumed to be the color of an old elliptical galaxy. Relations between mass-to-light ratio and color, based on SPS calculations of an evolved single burst by \citet{Bruzual2003} were then used to convert the observed cluster luminosities to stellar masses. The stellar masses were in turn used to normalise the SN Ia rates per unit mass. To obtain more direct and consistent estimates of the rates per unit mass that are implied by previous cluster SN Ia studies, we have re-assessed the previous rates, finding the \texttt{P\'egase.3} model $B$-band mass-to-light ratio for each cluster sample, and using it to translate the rates per unit luminosity to rates per unit mass. As previously, the \texttt{P\'egase.3} models assume a single star-formation burst,  while the delay time range allowed for each cluster corresponds to the time elapsed between formation redshifts in the range $z_f=3-4$ and the visibility-time-weighted average redshift of each cluster sample. In this uniform re-analysis or previous rates, we generally find higher mass-to-light ratios than previously assumed, and as a result the rates per unit mass for the low-$z$ clusters are about 30\% lower than estimated in the original publications.

Table~\ref{table:rates} summarises the updated SN Ia rates based on the single-burst assumption, for the \citetalias{Friedmann2018} sample, for its two subsamples, and for the previous measurements compiled by \cite{Maoz2017}. For the previous measurements, the SN Ia rates per unit $B$-band luminosity are reproduced, unchanged, from table 6 of \citetalias{Friedmann2018}, and the updated SN rates per unit formed stellar mass are based on the revised \texttt{P\'egase.3} mass-to-light ratios, as described above.

\begin{table}
	\caption{Revised SN Ia rates.}
	\begin{tabular*}{\linewidth}{l@{\extracolsep{\fill}}c@{\extracolsep{\fill}}c@{\extracolsep{\fill}}c@{\extracolsep{\fill}}c}
		\hline
		\hline
		\noalign{\vskip 0.5mm} 
        Source & $z$ & $t$ & $R_{\rm Ia,L_B}$ & $R_{\rm Ia, M}$\\
        & & [Gyr] & $\rm  [10^{-12}yr^{-1}L_{B,\odot}^{-1}]$ & $\rm  [10^{-14}yr^{-1}M_{\odot}^{-1}]$\\
        (1) & (2) & (3) & (4) & (5)\\
		\noalign{\vskip 0.5mm} 
		\hline
		\noalign{\vskip 0.5mm} 
		\multicolumn{5}{c}{\citetalias{Friedmann2018} sample, single redshift bin}\\
		\hline
		\noalign{\vskip 0.5mm} 
		FM18 & $1.35$ & $2.5\pm 0.3$ & $0.64\pm 0.19$ & $24.6\pm 7.1 $\\
		\noalign{\vskip 0.5mm} 
		\hline
		\noalign{\vskip 0.5mm} 
		\multicolumn{5}{c}{\citetalias{Friedmann2018} sample, two redshift bins}\\
		\hline
		\noalign{\vskip 0.5mm} 
		FM18 & $1.58$ & $2.1\pm 0.3$ & $0.74\pm 0.31$ & $37.0 \pm 15.5$\\
		\noalign{\vskip 0.5mm} 
		FM18 & $1.25$ & $3.0\pm 0.3$ & $0.57 \pm 0.20$ & $19.5 \pm 6.8$\\
		\noalign{\vskip 0.5mm} 
		\hline
		\noalign{\vskip 0.5mm} 
		\multicolumn{5}{c}{Previous measurements}\\	
		\noalign{\vskip 0.5mm} 
		\hline
		\noalign{\vskip 0.8mm} 	
		B12 & $1.12$ & $3.5 \pm 0.3$ & $0.50_{-0.28}^{+0.33}$ & $14.5_{-8.8}^{+10.3}$
		\\
		\noalign{\vskip 0.8mm} 
        GY02 & $0.90$ & $4.4 \pm 0.3$ & $0.80_{-0.52}^{+1.06}$ & $18.1_{-12.3}^{+24.1}$ 
        \\
        \noalign{\vskip 0.8mm} 
        S10 & $0.60$ & $5.9 \pm 0.3$ & $0.35_{-0.26}^{+0.31}$ & $5.8_{-4.8}^{+5.5}$
        \\
        \noalign{\vskip 0.8mm} 
        G08 & $0.46$ & $6.9 \pm 0.3$ & $0.31_{-0.16}^{+0.51}$ & $4.5_{-2.4}^{+7.4}$
        \\
        \noalign{\vskip 0.8mm} 
        GY02 & $0.25$ & $8.7 \pm 0.3$ & $0.39_{-0.32}^{+0.90}$ & $4.6_{-3.8}^{+10.8}$
        \\
        \noalign{\vskip 0.8mm} 
        D10 & $0.23$ & $8.9 \pm 0.3$ & $0.33_{-0.08}^{+0.09}$ & $3.9_{-1.1}^{+1.3}$
        \\
        \noalign{\vskip 0.8mm} 
        S07 & $0.15$ & $9.8 \pm 0.3$ & $0.36_{-0.16}^{+0.24}$ & $3.8_{-1.8}^{+2.6}$
        \\
        \noalign{\vskip 0.8mm} 
        D10 & $0.08$ & $10.6 \pm 0.3$ & $0.23_{-0.08}^{+0.11}$ & $2.3_{-0.9}^{+1.2}$
        \\
        \noalign{\vskip 0.8mm} 
        M08 & $0.02$ & $11.4 \pm 0.3$ & $0.28_{-0.08}^{+0.11}$ & $2.6_{-0.9}^{+1.1}$
        \\
        \noalign{\vskip 0.8mm} 
        \hline
	\end{tabular*}
	\begin{minipage}{1\linewidth}
	\textbf{Notes.} 
	(1) References for the different measurements: \citetalias{Friedmann2018}; B12 -- \citet{Barbary2012}; GY02 -- \citet{Gal-Yam2002}; S10 -- \citet{Sharon2010}; G08 -- \citet{Graham2008}; D10 -- \citet{Dilday2010}; S07 -- \citet{Sharon2007}; M08 -- \citet{Mannucci2008}. 
    (2) Visibility-time-weighted average redshift of each cluster sample.
    (3) Delay time corresponding to a formation redshift between $z_f=3-4$, given the the visibility-time-weighted average redshift.
    (4) SN Ia rate per unit rest-frame $B$-band stellar luminosity, derived as explained in section~\ref{subsection:revised_SNIa} for the \citetalias{Friedmann2018} sample, and as indicated in table 6 of \citetalias{Friedmann2018} for the previous measurements. 
    (5) SN Ia rate per unit formed stellar mass, derived as explained in section~\ref{subsection:revised_SNIa}. The average $B$-band mass-to-light ratios from \texttt{P\'egase.3} models can be recovered by dividing columns (4) by (5). 
    \end{minipage}
	\label{table:rates}
\end{table}

\section{Modeling}
\label{section:modeling}

With the revised fluxes and SNe Ia numbers for the \FM18 sample and the revised lower-$z$ SNe Ia rates in hand, we now attempt to constrain the SN Ia DTD in cluster environments in the presence of extended star formation histories (SFHs) for the monitored \FM18 clusters. To this end, we calculate SPS models, and find the range of SFH parameters that are consistent with the observed spectral energy distributions (SEDs) of the clusters in the HST bands. Simultaneously, we find the allowed range of the parameters describing the DTD, by fitting the observed number of SNe Ia in each cluster in \FM18  to the predictions for a given SFH plus DTD combination, and further by taking into account previous measurements of the SN Ia rate per unit formed stellar mass in lower-$z$ cluster samples.

\subsection{Star-formation parameters and SED fitting}
\label{subsection:Pegase}

We constrain the range of SFH parameters, that are consistent with the observed HST stellar fluxes for each cluster,  by producing model SEDs with the \texttt{Pégase.3} SPS code. 
To investigate a broad range of possible SFHs for the clusters, yet without introducing too many free parameters, we assume, for each cluster, an exponentially decaying star formation rate (SFR),
\be
\label{eq:SFH}
\psi(t)=\frac{M_0}{\tau} \exp(-t/\tau) ,
\ee
having three parameters: 
(i) the initial star-formation redshift, $z_f$, which is equivalently represented with a parameter $t$, the cosmic time elapsed between $z_f$ ($t=0)$ and $z$ at which the cluster is observed;
(ii) $\tau$, the characteristic time of the exponential; and
(iii) $M_0$ the asymptotically (at infinite time) formed stellar mass. 
The limit $\tau=0$ corresponds to an instantaneous burst of star formation, with $\psi(t) = M_0~\delta(t)$ where $\delta$ is the Dirac delta function. 
In the SPS models, as before, we adopt an unevolving metallicity of $Z=0.020$, based on the close-to-solar abundances generally observed in the massive early-type galaxies that make up the bulk of the stellar mass in galaxy clusters \citep[e.g.][]{Conroy2014}, a \citet{Kroupa2001} IMF between 0.1 and $100~M_\odot$, and Padova+AGB stellar evolution tracks.
A fourth modeling parameter that we adopt, which relates the cluster's model SFH to its observed SED, is a rest-frame $V$-band dust extinction parameter $A_V$, based on the \citet{Cardelli1989} extinction curve with reddening parameter $R_V=3.1$, applied as a foreground dust screen to the synthesized galaxy light.  
Finally, we allow for a "late" SFH component in the form of a second exponential burst, beginning 100 Myr before the cluster-observation epoch, with a fixed exponential timescale of 100 Myr. This SFH model component opens up an option to account for the blue emission sometimes apparent in the cluster SEDs, and indeed in some individual cluster galaxies with indications of recent star formation. The choice of 100 Myr for both the delay and for the exponential timescale of this component allows for a significant effect, both on the light (this late stellar population is still young and forming) and on the SN Ia rate ($>40$~Myr have elapsed, so the DTD has "turned on", and the DTD is still near its peak SN Ia rate). 
This second burst of star formation is parametrised with $m=M_{0}^\prime/M_0$, where $M_0^\prime$ is the asymptotically formed stellar mass of the late burst. The late burst is assumed to have the same constant solar metallicity as the main burst, and to be unaffected by dust. 
The assumed SFH of each cluster is thus described by five parameters: $t$, $\tau$, $M_0$, $A_V$, and $m$, with $M_0$ being essentially a scaling factor that matches the calculated fluxes to the observed ones. 

For each set of parameters, we use \texttt{P\'egase.3} to derive a model spectrum for a unit stellar mass. For each cluster in the observed sample, we shift the model spectrum to the cluster redshift, and convert the model's luminosity density per unit wavelength interval, $L_\lambda$, to a flux density $f_\lambda=L_\lambda/[4\pi D_L^2 (1+z)]$, where $D_L$ is a cluster's luminosity distance, assuming standard cosmological parameters ($\Omega_m=0.3$ , $\Omega_\Lambda=0.7$, and $H_0=70~\rm km s^{-1} Mpc^{-1}$). The spectrum is then folded through the HST+filter system bandpasses in F105W, F140W, and F160W, to obtain the model band-averaged flux in each observed band, which can be compared to the measured fluxes.

We use $\chi^2$ as a figure of merit to gauge the agreement of a SFH model with the observed fluxes of a particular cluster:
\be
\label{eq:chi2}
\chi^2 = \sum\limits_i \frac{\left(f_{i}-M_0 f_{i, \rm model}\right)^2}{{\rm d}f_{i}^2}, 
\ee
where $f_{i}$ is a measured integrated-galaxy-cluster flux in the $i^{\rm th}$ band, ${\rm d}f_{i}$ is its associated uncertainty (see Table~\ref{table:fluxes}), $f_{i,\rm model}$ is the predicted flux \textit{per unit formed stellar mass} for the given SFH model, and the best-fit asymptotically formed stellar mass is 
\be
\label{eq:mstar}
M_0 = \frac{\sum\limits_i \frac{f_{i} f_{i,\rm model}}{{\rm d}f_{i}^2}}{\sum\limits_i \frac{f_{i,\rm model}^2}{{\rm d}f_{i}^2}}, 
\ee
i.e., the weighted average of the ratios $f_{i}/f_{i,\rm model}$, given the observational uncertainties.

For each of the parameters of the SFH modeling, we explore the following ranges. The delay-time range corresponds to formation redshifts in the range $z_f=3-4$ (see \S 1). 
We restrict the SFR timescale $\tau$ to the range $0-1.2$~Gyr. This is based on the finding by \cite{Salvador-Rusinol2020} that $<0.5$\% of the stellar mass of massive early-type galaxies at $z\sim0.4$ was formed in the preceding 2~Gyr. For a galaxy with an exponentially declining SFR, formed at $z_f=4$, creation of  $<0.5$\% of its mass between lookback times of 6 Gyr and 4 Gyr (i.e. $z\approx 0.4$ and 2 Gyr earlier), requires $\tau<1.2$~Gyr. (A stricter constraint would result if formation were assumed at $z_f=3$). This range in $\tau$ thus assures that our assumed SFHs for the clusters are consistent with the sensitive limits set by \cite{Salvador-Rusinol2020} on the presence of any young stellar populations in early-type galaxies.
We limit the amount of dust extinction to $A_V < 1.5$~mag; the modeling of the individual SN Ia light curves in \citetalias{Friedmann2018}, as well as SPS models fit to the spectra of the individual SN host galaxies in \citetalias{Williams2020}, both yield typical extinctions below this value. Early-type galaxies, as a rule, do not reveal evidence of much dust. 
The parameter $m$, embodying the effects of late-time star formation, is allowed to vary between $10^{-3}$ and $1$.

\begin{figure*}
	\includegraphics[height=0.4\textwidth,trim={1.3cm 0.3cm 1.cm 1.1cm},clip]{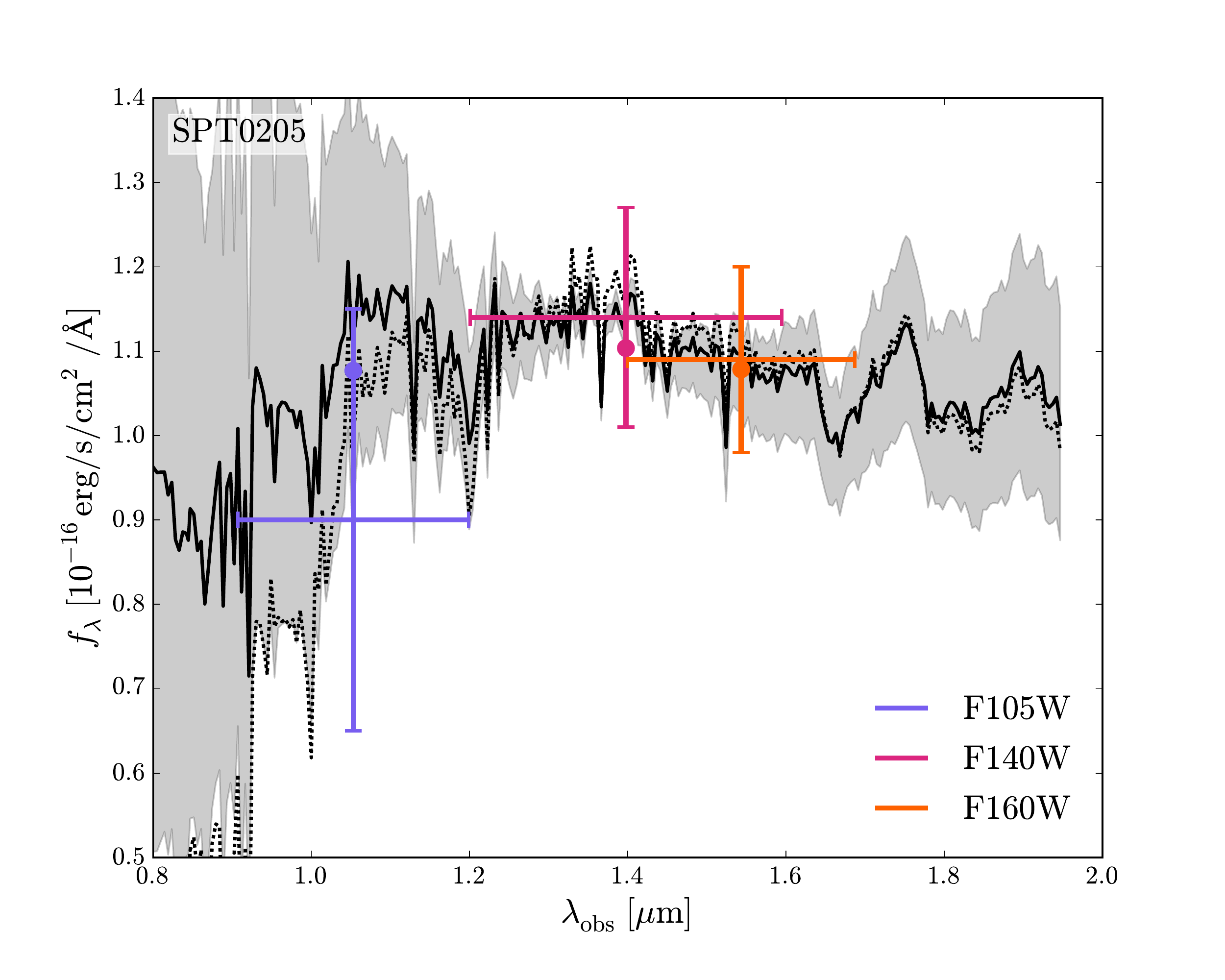}\hfill
	\includegraphics[height=0.4\textwidth,trim={1.3cm 0.3cm 1.cm 1.1cm},clip]{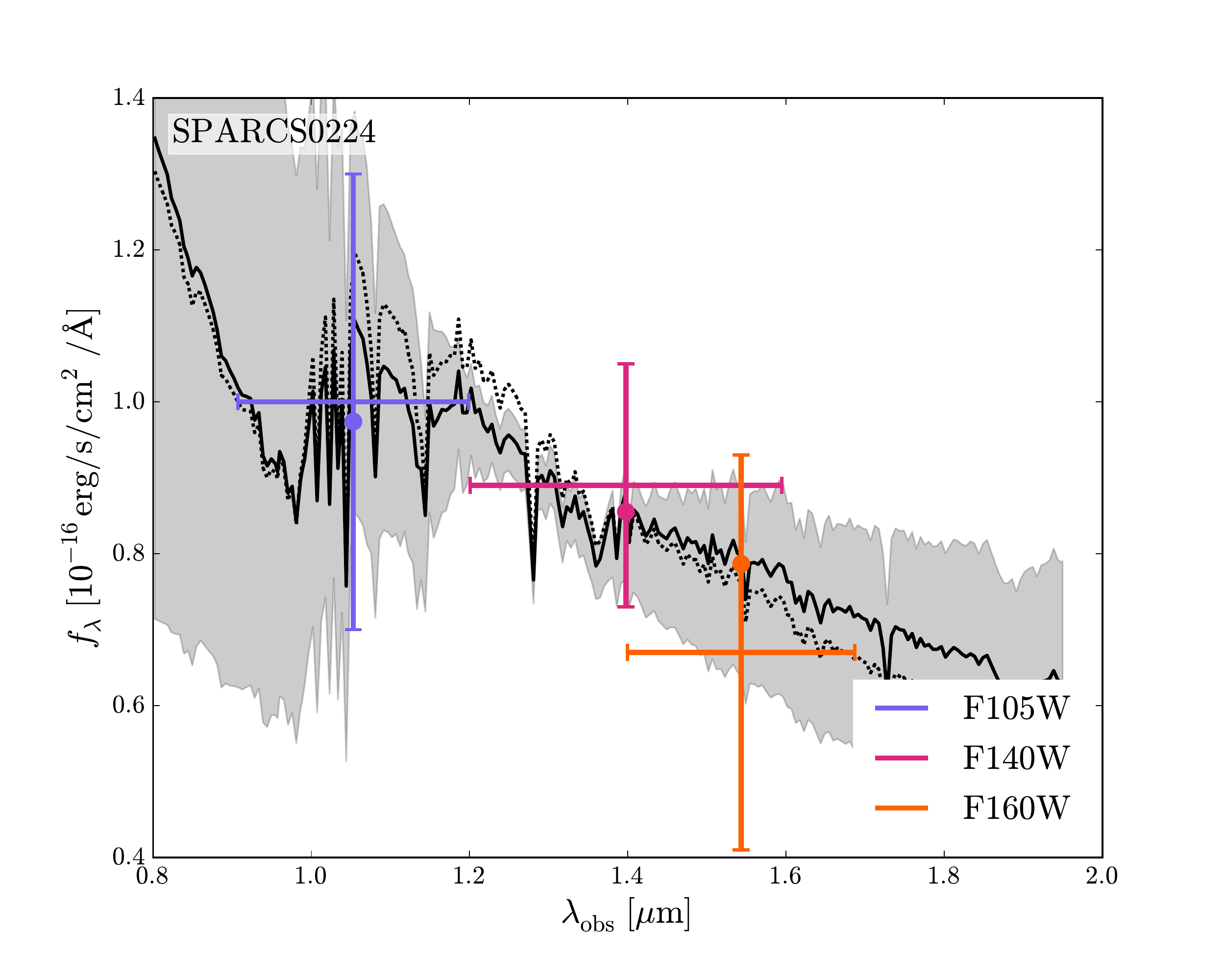}
	\\
	\includegraphics[height=0.4\textwidth,trim={1.3cm 0.3cm 1.cm 1.1cm},clip]{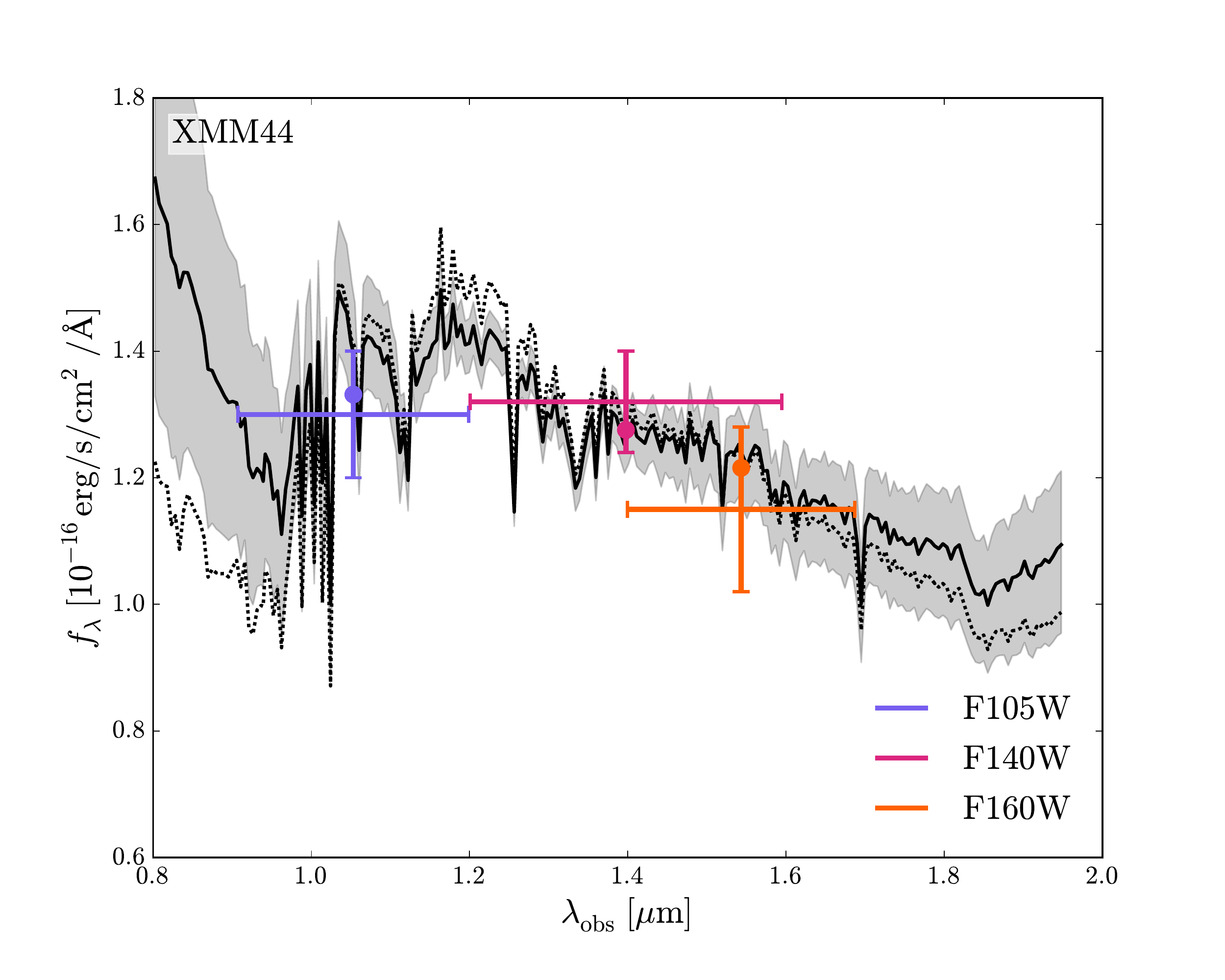}\hfill
	\includegraphics[height=0.4\textwidth,trim={1.3cm 0.3cm 1.cm 1.1cm},clip]{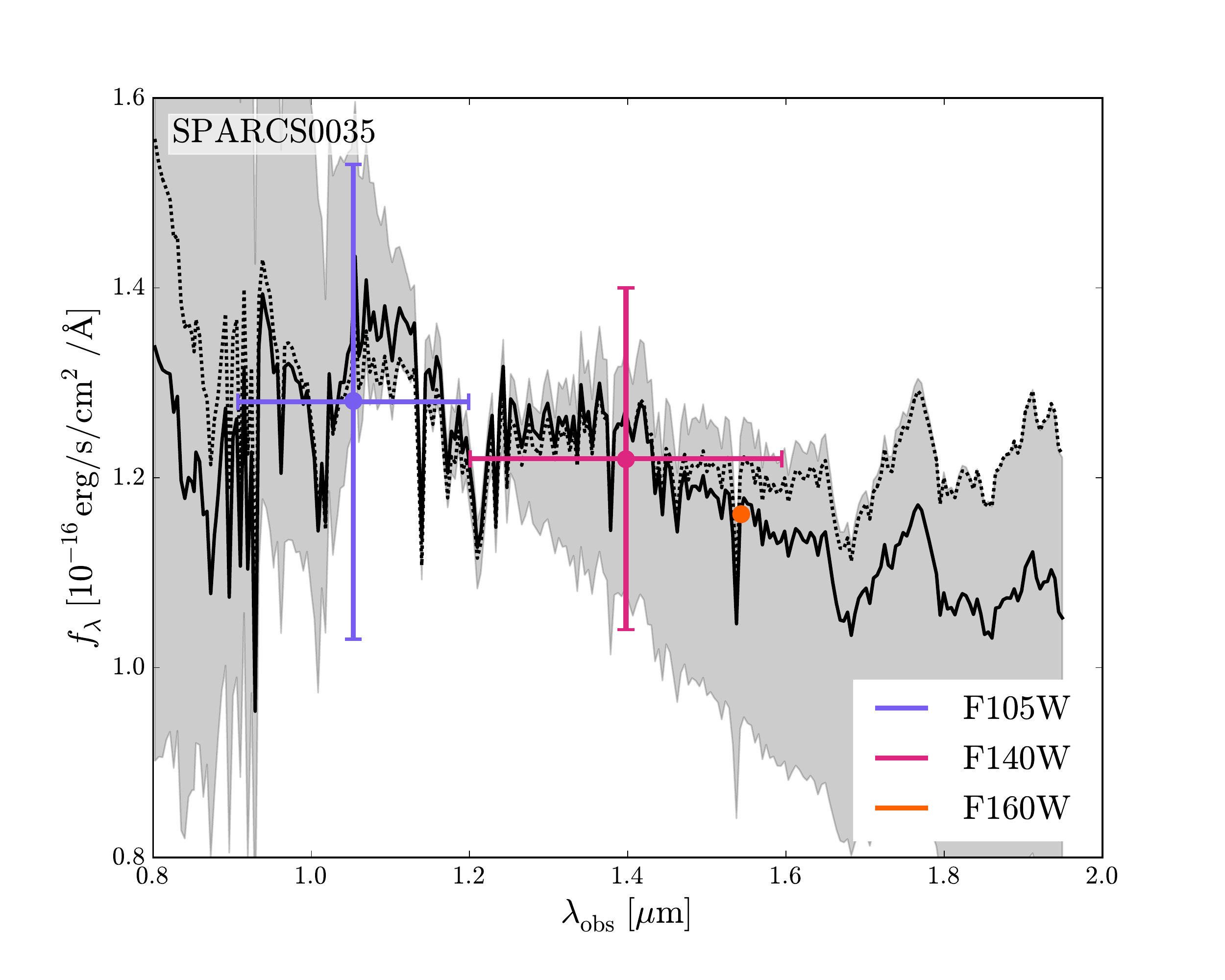}
	\vspace{-0.2cm}
	\caption{
	Observed fluxes and \texttt{P\'egase.3} model spectra for four of the \protect\FM18 clusters. The dotted spectrum corresponds to the best-fit (lowest-$\chi^2$) model, the solid spectrum to the median of the allowed models, and the gray area to the $1\sigma$ range.  Observed fluxes and uncertainties in the F105W, F140W, and F160W HST bands are indicated by vertical error bars. Horizontal error bars denote the filter bandpasses. Predicted fluxes, from folding the median model spectrum through the bandpass responses, are marked with filled circles. 
	}
	\label{fig:chi2_example}
\end{figure*}

To illustrate the SED-fitting process, Fig.~\ref{fig:chi2_example} shows the best-fit spectrum and the range of allowed spectra for four of the \FM18 clusters. While the parameter $m$ is generally limited by the fits to small values between $10^{-3}$ and $10^{-2}$, the other SFH parameters are weakly constrained by the cluster photometry alone, and thus they can span much of their allowed range.

\subsection{DTD parameters and modeling of SN Ia numbers }
\label{subsection:NSN_model}

In line with previous work, we will assume a DTD having a power-law form, with units of SN Ia rate per formed stellar mass,
\be
\label{eq:DTD}
{\rm DTD} (t) = R_1 \left( \frac{t}{\rm Gyr}\right)^\alpha, 
\ee
where $R_1$ is the DTD rate at $t=1$~Gyr.
The SN Ia rate at time $t$ in a galaxy cluster with a given SFH, $\psi(t)$, is the convolution of the SFH with the DTD, 
\be
\label{eq:RSNIa_model}
R_{\rm Ia, model} (t) = \int_{t_{\rm min}}^t \psi(t-t^\prime) \times  {\rm DTD} (t^\prime) dt^\prime. 
\ee
As already noted, $t_{\rm min}=40~\rm Myr$ is the stellar evolution time until the appearance of the first white dwarfs in a stellar population. 
For the exponentially decaying SFH of Eq.~(\ref{eq:SFH}), Eq.~(\ref{eq:RSNIa_model}) becomes
\be
\label{eq:RSNIa_model_exp}
R_{\rm Ia, model} (t) = \frac{M_0 R_1}{\tau} \int_{t_{\rm min}}^t \left(\frac{t^\prime}{\rm Gyr}\right)^\alpha \exp\left(\frac{t^\prime-t}{\tau}\right)  dt^\prime, 
\ee
which simplifies to $M_0 R_1 (t/{\rm Gyr})^\alpha$ for an instantaneous burst. For our SFH models that include a second burst, 100 Myr before the redshift  of observation, we add to the SN Ia rate from the main, initial burst, the SN Ia rate induced by this second burst.

The rest-frame monitoring times and the experiment's detection-completeness factors for SNe Ia that reached maximum light before, during, and after each HST observing "season" for each cluster, were derived based on simulations in \citetalias{Friedmann2018}, and are reproduced in Table~\ref{table:fluxes}. The number of detected SNe Ia predicted for the experiment in a given cluster by a given SFH model and a particular DTD model is 
\be
N_{\rm Ia, model} = R_{\rm Ia, model}(t) \times \sum\limits_i \tau_{{\rm vis},i} ~\eta_i ,
\ee
where the sum is both over the observing seasons (if more than one) and the periods before, during and after each season. For each cluster, the $\tau_{{\rm vis},i}$ are the rest-frame monitoring times, i.e., the cluster rest-frame interval of actual monitoring, a $40/(1+z)$-day interval before the monitoring period, and a $20/(1+z)$-day interval after the monitoring period, and $\eta_{0}$, $\eta_{-40}$, and $\eta_{+20}$ are the corresponding detection-completeness factors.
This expected number of SN Ia can be compared to the actual number observed in each cluster, constraining
both the SFH model and the DTD parameters.

\begin{figure}
    \centering
	\includegraphics[height=0.4\textwidth,trim={0.2cm 0.5cm 0.cm -0.3cm},clip]{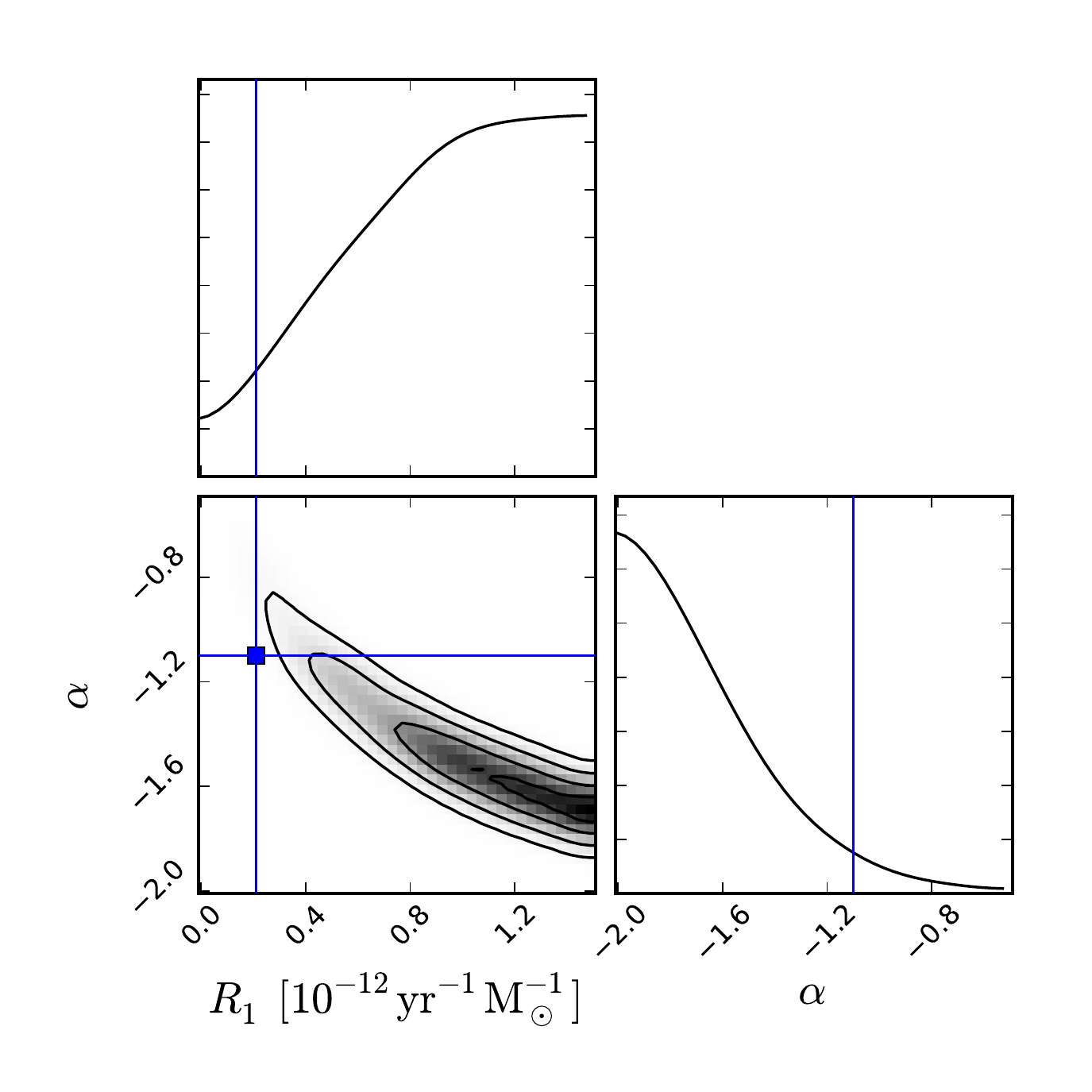}
	\vspace{-0.1cm}
	\caption{
	Prior distribution of the DTD parameters $R_1$ and $\alpha$, based on the lower-$z$ cluster sample SN rate measurements compiled by \FM18 and revised in section~\ref{subsection:revised_SNIa}. The prior is assumed to follow a chi-square probability with one degree of freedom, as indicated in Eq.~(\ref{eq:prior}). For reference, the blue square highlighted with a solid cross corresponds to the field-galaxy DTD parameters from \protect\citet{Maoz2017}: $R_1=(0.21\pm 0.02) \times 10^{-12}~\rm yr^{-1}M_\odot^{-1}$ and $\alpha=-1.07\pm 0.09$. Contours correspond to 0.5, 1, 1.5, and 2 $\sigma$ levels. The low-$z$ cluster SN rates alone do not constrain strongly the DTD, with a strong anti-correlation between its two parameters.
	}
	\label{fig:Plow}
\end{figure}

Since the low-$z$ cluster samples whose SN Ia rates were compiled in \FM18 are observed at relatively large delay times, for which the assumption of passive evolution after a brief burst at high $z$ is likely valid, their observed SN Ia rates per formed stellar mass (as revised in section~\ref{subsection:revised_SNIa}) are direct measures of the DTD at different delays. To obtain prior constraints on the DTD parameters imposed by these previous low-$z$ SN Ia rate measurements, we quantify the agreement of a given DTD model $(R_1,\alpha)$ with the measurements through
\be
\label{eq:chi2_lowz}
\chi_{\rm low}^2 = \sum_j \left(\frac{R_{{\rm Ia,M},j}-{\rm DTD}(t_{j}|R_1,\alpha)}{dR_{{\rm Ia,M},j}}\right)^2, 
\ee 
where, for the $j^{\rm th}$ low-$z$ cluster sample, $t_j$, $R_{{\rm Ia,M},j}$ and $dR_{{\rm Ia,M},j}$ are respectively the delay time, the rate per unit formed stellar mass, and its uncertainty, as indicated in table~\ref{table:rates}.
When $R_{{\rm Ia,M},j}<{\rm DTD}(t_{j})$, we use the positive uncertainty; when $R_{{\rm Ia,M},j}>{\rm DTD}(t_{j})$, the negative one.
The prior distribution of $R_1$ and $\alpha$, based on the low-$z$ cluster SN Ia rates, is therefore the chi-square probability with one degree of freedom
\be
\label{eq:prior}
\mathcal{P}_{\rm low}(R_1,\alpha) = \frac{\chi_{{\rm low}}^{-1} e^{- \chi_{{\rm low}}^2/2}}{2^{\nicefrac{1}{2}} \Gamma(\nicefrac{1}{2})}, 
\ee
which is shown in Fig.~\ref{fig:Plow}. As noted in \citet{Maoz2010}, where the SN Ia rates from these low-$z$ clusters were already analysed, these low-$z$ rates alone do not have sufficient leverage in terms of redshift range (translating to delay-time range) in order to individually constrain well the two DTD parameters, which are strongly anti-correlated in the fit. However, as we will see below, when combined, as a prior, with the high-$z$ (short-delay) SN Ia measurements of \FM18, the low-$z$ measurements do significantly tighten the constraints on the DTD.

\subsection{Simultaneous SFH and DTD model fits to the data}
\label{subsection:MCMC}

We use Bayesian inference to derive posterior probabilities for the SFH and DTD parameters, i.e., 
\be
\mathcal{P}(\theta | D) \propto \mathcal{P}(D|\theta) \times  \mathcal{P}(\theta) 
\ee
where $\theta$ indicates the parameters, $D$ the data, $\mathcal{P}(\theta)$ the prior, and $\mathcal{P}(D|\theta)$ the likelihood. 
For describing the SFH of each of the 12 \citetalias{Friedmann2018} clusters, each with four SFH parameters ($t$, $\tau$, $m$, and $A_V$), $\theta$ is a vector of $12 \times 4 +2 = 50$ elements with the two parameters of the universal DTD ($R_1$, $\alpha$). 
$D$ is a vector of $2\times 32+12+2\times12+4\times 18=172$ elements: 32 fluxes in different bands and their uncertainties, 12 redshifts, 12 counts of likely SNe Ia, 12 counts of possible SNe Ia, and 18 monitoring seasons and their detection-completeness factors.

The prior is assumed uniform for all SFH parameters, with the following boundaries: $t$ corresponds to a formation redshift between $z_f=3-4$, $\tau$ is between 0 (instantaneous burst) and 1.2 Gyr, $A_V$ between 0 and 1.5 mag, $\log m$ is between $-3$ and $0$ (see last paragraph of Section~\ref{subsection:Pegase}, for a discussion of the boundaries chosen for $\tau$ and $A_V$). 
The DTD parameters $\alpha$ and $R_1$ are assumed to be, respectively, between $-5$ and 0 and between 0.01 and $5\times 10^{-12}~\rm yr^{-1}M_\odot^{-1}$. When considering only the \FM18 clusters and their SNe Ia, we assume uniform priors for the DTD parameters. To take into account the constraints on the DTD parameters stemming from the low-$z$ cluster measurements, we assume the prior $\mathcal{P}(\theta)=\mathcal{P}_{\rm low}(R_1,\alpha)$ from Eq.~(\ref{eq:prior}).

The logarithm of the likelihood is the sum of probabilities
\be
\label{eq:likelihood}
\log \mathcal{P}(D|\theta) = \sum_i \log \mathcal{P}_{\rm SFH, i} + \sum_i \log \mathcal{P}_{\rm FM18, i}, 
\ee
where
\be
\mathcal{P}_{{\rm SFH},i} = \frac{\chi_i^{-1} e^{- \chi_i^2/2}}{2^{\nicefrac{1}{2}} \Gamma(\nicefrac{1}{2})}
\ee
is the chi-square probability with one degree of freedom associated with the $\chi^2$ of Eq.~(\ref{eq:chi2}) for cluster $i$, and 
\be
\mathcal{P}_{{\rm FM18}, i} = \frac{\left(N_{{\rm Ia, model,} i}\right)^{N_{{\rm Ia, obs,} i}} e^{-N_{{\rm Ia, model,} i}} }{\Gamma(N_{{\rm Ia, obs,} i}+1)} 
\ee
is the Poisson probability of an observed SN Ia count $N_{{\rm Ia, obs,}i}$ in cluster $i$ given a count predicted by a model with particular SFH and DTD parameters. 
The formulation of the Poisson probability with the gamma function $\Gamma$ permits considering non-integer "observed" counts of SNe---
as in \citetalias{Friedmann2018}, the observed count of SNe Ia in each of the  clusters is taken as
\be
\label{eq:Nobs}
N_{{\rm Ia, obs}} = N_{\rm Ia}+0.5 N_{\rm Ia?}, 
\ee
where $N_{\rm Ia}$ and $N_{\rm Ia?}$ are, respectively, the number of likely and possible cluster SNe Ia, as indicated in Table~\ref{table:fluxes}. The two sums of Eq.~(\ref{eq:likelihood}) are over the 12 \citetalias{Friedmann2018} high-$z$ clusters at the focus of the present work.

\label{subsection:setups}

To understand the results of the present analysis and any differences from previous studies of the cluster-environment SN Ia DTD,   
we consider four different setups, to gauge both the effect of assuming extended SFHs versus single instantaneous starbursts (as in previous studies) and the effect of considering the data for the lower-$z$ clusters: 

\noindent\textbf{$\bullet$ Setup 1.} Only the \citetalias{Friedmann2018} clusters and their SNe Ia are modeled, with their SFHs described by single short bursts. In this case, the model has 14 parameters: $t$ for each of the 12 clusters, $R_1$, and $\alpha$. We assume $\tau=0$, no extinction ($A_V=0$), and no late burst ($m=0$). Uniform priors are assumed for $t$, $R_1$, and $\alpha$.

\noindent\textbf{$\bullet$ Setup 2.} Only the \citetalias{Friedmann2018} clusters, but allowing for extended SFHs in the clusters. In this case, the model has 50 parameters: $t$, $\tau$, $A_V$, and $m$ for each of the 12 clusters, $R_1$, and $\alpha$. 
Uniform priors are again assumed for all parameters.

\noindent\textbf{$\bullet$ Setup 3.} The \citetalias{Friedmann2018} clusters are considered, together with low-$z$ constraints on the two DTD parameters, but again with all SFHs described by single short bursts. As for setup 1, the model has 14 parameters: $t$ for each of the 12 clusters, $R_1$, and $\alpha$.
We assume $\tau=0$, no extinction ($A_V=0$), no second burst ($m=0$). The  
prior for the DTD parameters, based on low-$z$ cluster rate measurements, is $\mathcal{P}_{\rm low}(R_1,\alpha)$ from Eq.~(\ref{eq:prior}). A uniform prior is assumed for $t$.

\noindent\textbf{$\bullet$ Setup 4.} We model the \citetalias{Friedmann2018} clusters, permitting extended cluster SFHs, {\it and} considering the constraints on the two DTD parameters from the low-$z$ cluster samples.
As for setup 2, the model has 50 parameters: 
$t$, $\tau$, $A_V$, and $m$ for each of the 12 clusters, $R_1$, and $\alpha$. 
While uniform priors are assumed for the SFH parameters, the prior for the DTD parameters is $\mathcal{P}_{\rm low}(R_1,\alpha)$ from Eq.~(\ref{eq:prior}).
Amongst the four setups, this setup makes maximal use of the observational data to constrain the DTD while allowing a wide range of SFHs. We base our conclusions on setup 4.

With the definition of the likelihood $\mathcal{P}(D|\theta)$ and the prior $\mathcal{P}(\theta)$, we construct the posterior probability $\mathcal{P}(\theta|D)$ using the Monte Carlo Markov Chain (MCMC) \texttt{python} implementation \texttt{emcee} by \cite{Foreman-Mackey2013}. 
We have calculated the four different MCMC setups outlined above, to derive the posterior probabilities of the parameters describing the cluster SFHs and, more importantly, the DTD parameters. In the \texttt{emcee} implementataion of the MCMC, we use 500 walkers.
We monitor the convergence of the chains by following the walker positions, their medians and standard deviations, the evolution of the posterior probability, as well as the autocorrelation time. The integrated autocorrelation times are $\sim$3,000 iterations for setups 1 and 3 and $\gtrsim$10,000 for setups 2 and 4. We let the chains reach 100,000 iterations for the four setups, and estimate the posterior distributions over the last 10,000 iterations.

\section{Results}
\label{section:results}

\begin{figure*}
	\includegraphics[width=0.495\linewidth,trim={.1cm 0.1cm 1.6cm 0.9cm},clip]{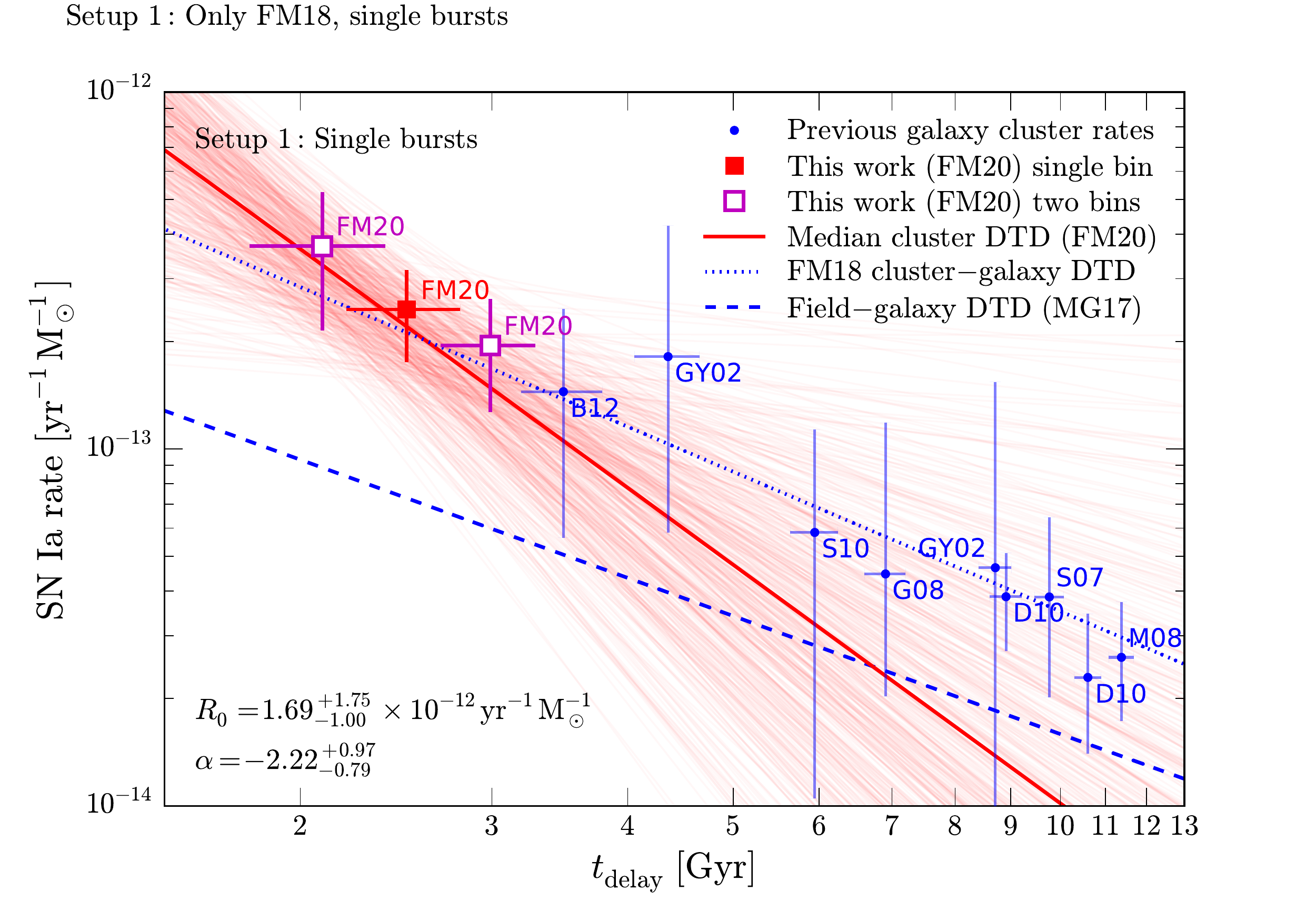}
	\hfill
	\includegraphics[width=0.495\linewidth,trim={.1cm 0.1cm 1.6cm 0.9cm},clip]{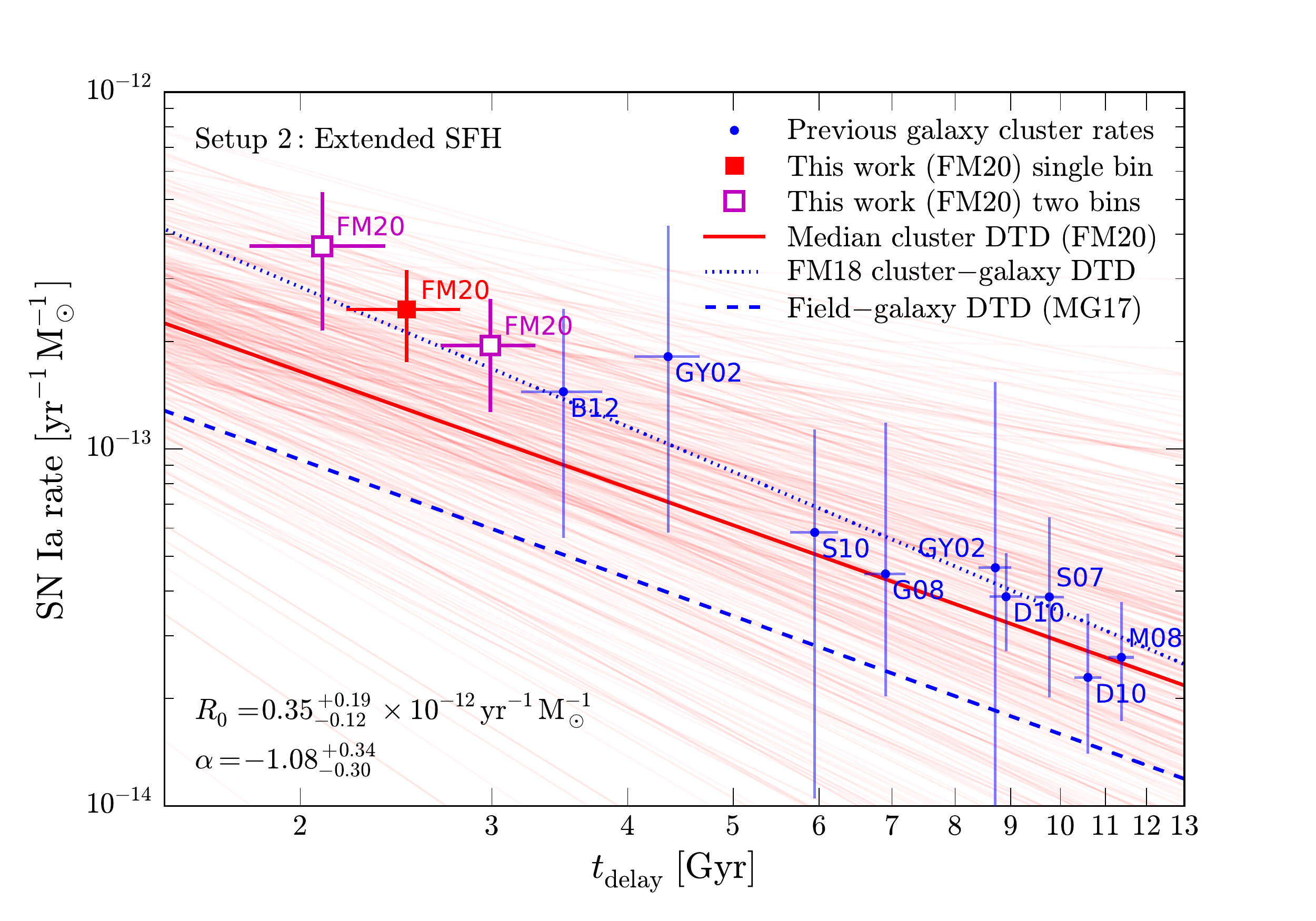}
    \\
	\includegraphics[width=0.495\linewidth,trim={.1cm 0.1cm 1.6cm 0.9cm},clip]{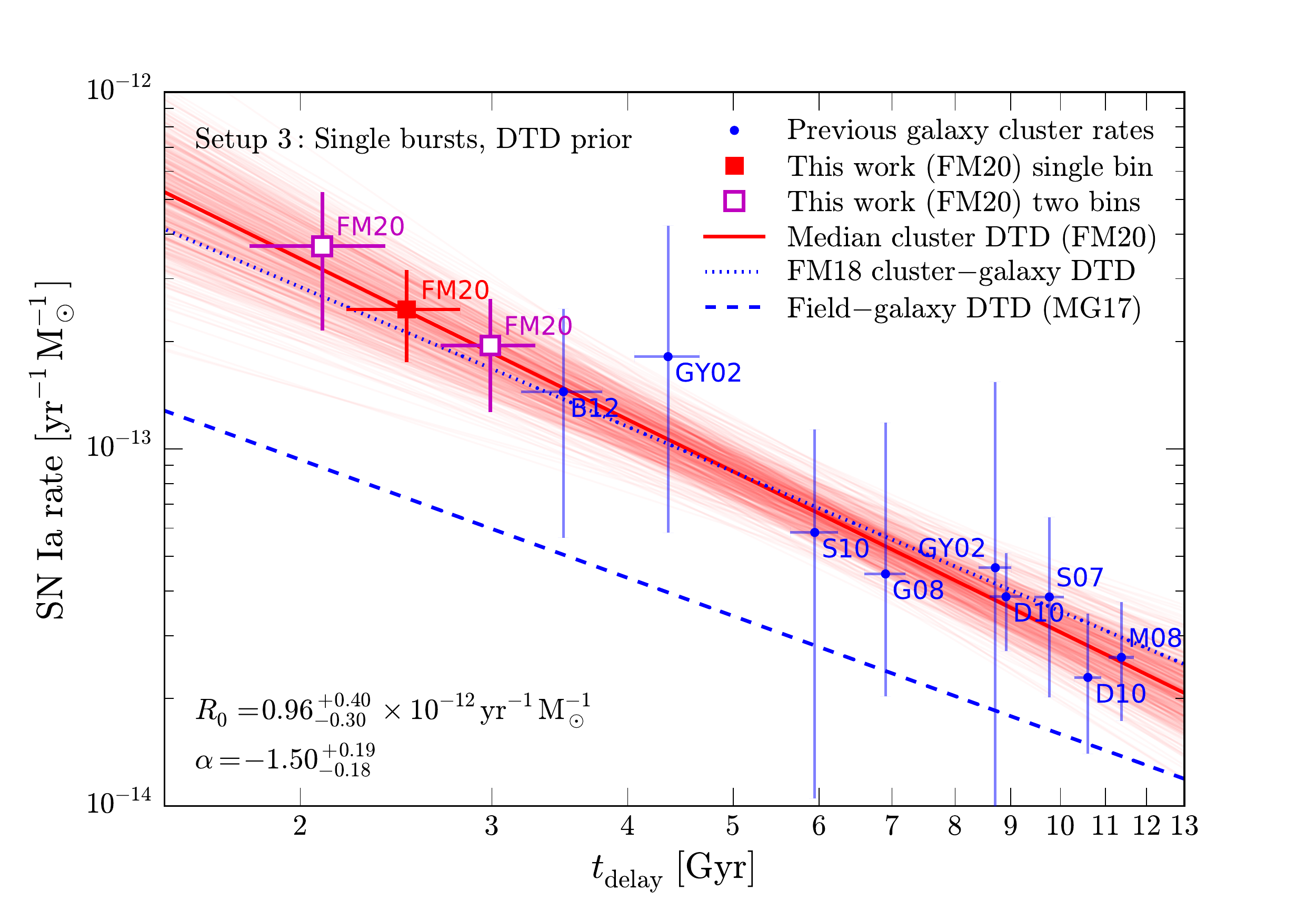}
	\hfill
	\frame{\includegraphics[width=0.495\linewidth,trim={.1cm 0.1cm 1.6cm 0.9cm},clip]{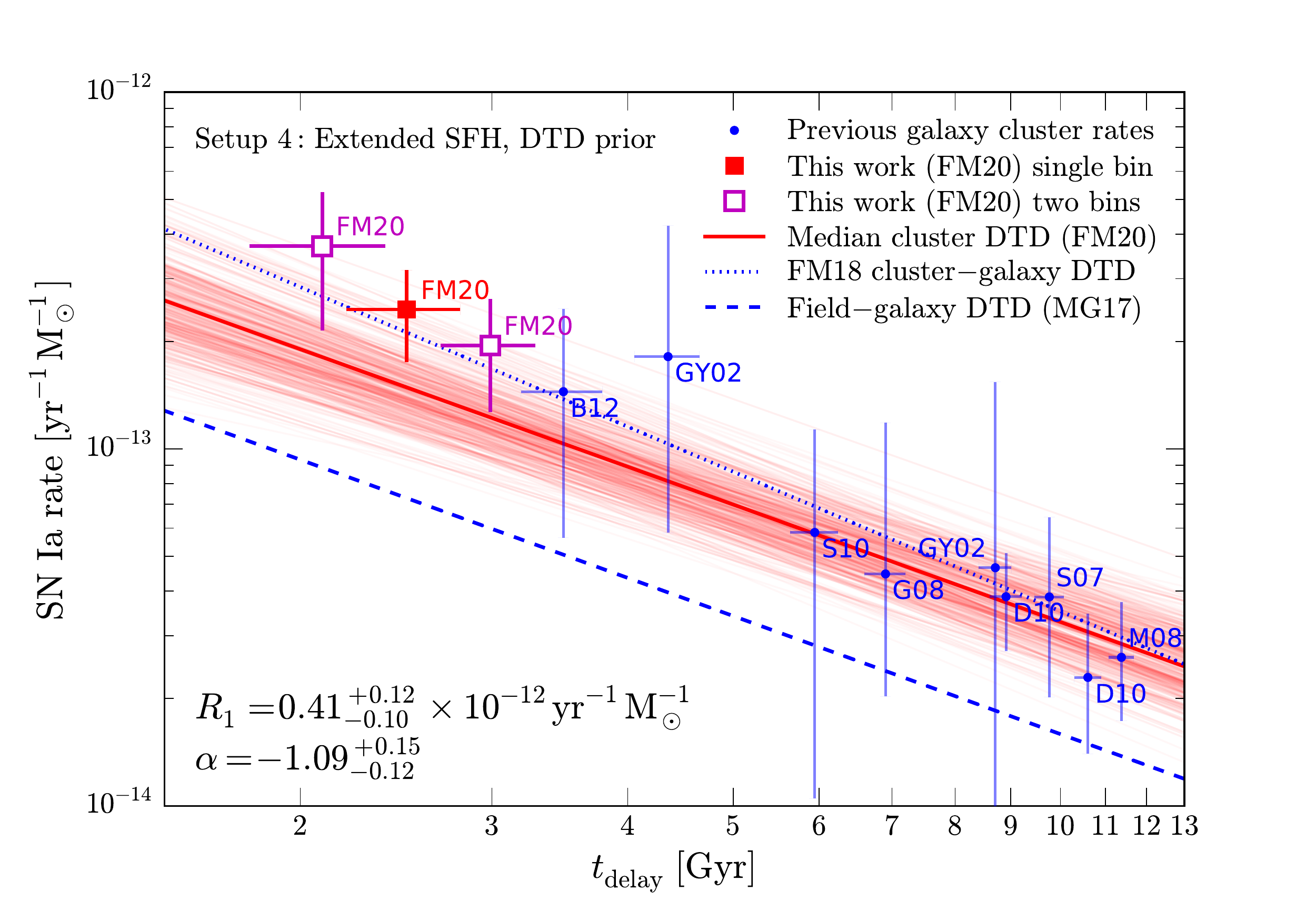}}
	\caption{
	Cluster-environment power-law DTD models emerging from our MCMC Bayesian inference, given the four different modeling setups (see section~\ref{subsection:setups}). 
	Thin red lines sample the DTD parameters resulting from the MCMC, while the thick red line shows the DTD with the median parameters in the converged MCMC calculation. The cluster-environment best-fit DTD from \citetalias{Friedmann2018} is plotted as a dotted blue line, and the field-galaxy DTD from \protect\cite{Maoz2017} as a dashed blue line. Data points, shown here for reference only (i.e. these data points are not being fit), are the SN Ia rates per unit formed stellar mass, in the \FM18 cluster samples and in previously studied, lower-$z$ cluster samples, all with values as revised in the present paper and presented in section \ref{subsection:revised_SNIa}. These plotted rates are based on the assumption that the stars in all the clusters were formed in a single burst between $z=3$ and 4, an assumption that we forego in the model setups shown in the right-hand panels. Comparison of the two bottom panels shows that allowing, in the \FM18 clusters, for extended SFHs (setup 4) that are consistent with the observed luminosities and colors, leads to a cluster SN Ia DTD (bottom right panel) that is shallower and lower-normalised than the DTD with the single-burst assumption (setup 3, bottom-left panel). The setup-4 cluster DTD, however, is still significantly higher-normalised than the field-SN DTD, albeit only by a factor of 2-3.}
	\label{fig:DTD}
\end{figure*}

\begin{figure*}
    \centering
	\includegraphics[height=0.4\textwidth,trim={0.2cm 0.5cm 0.cm -0.3cm},clip]{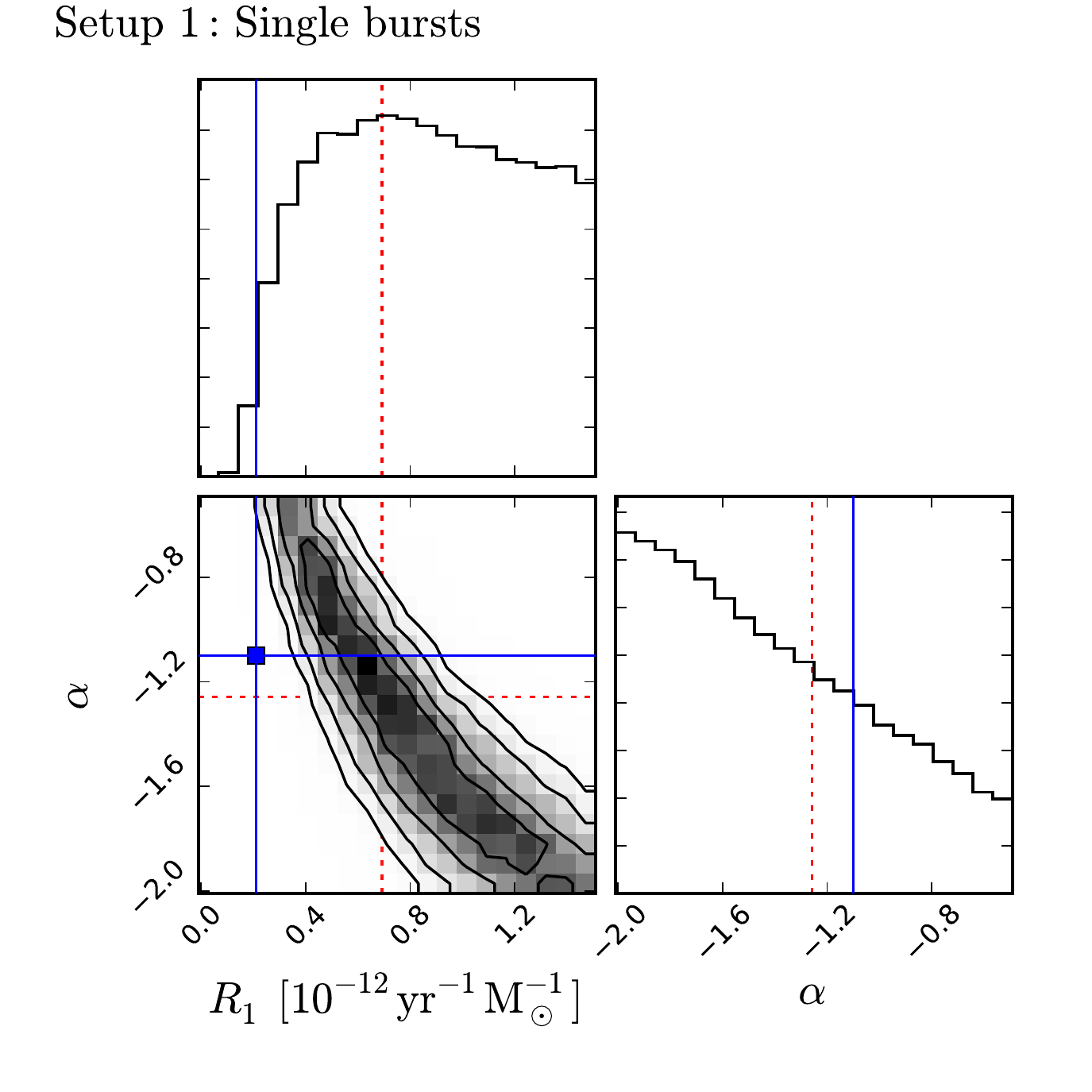}
	\hspace{0.6cm}
	\includegraphics[height=0.4\textwidth,trim={0.2cm 0.5cm 0.cm -0.3cm},clip]{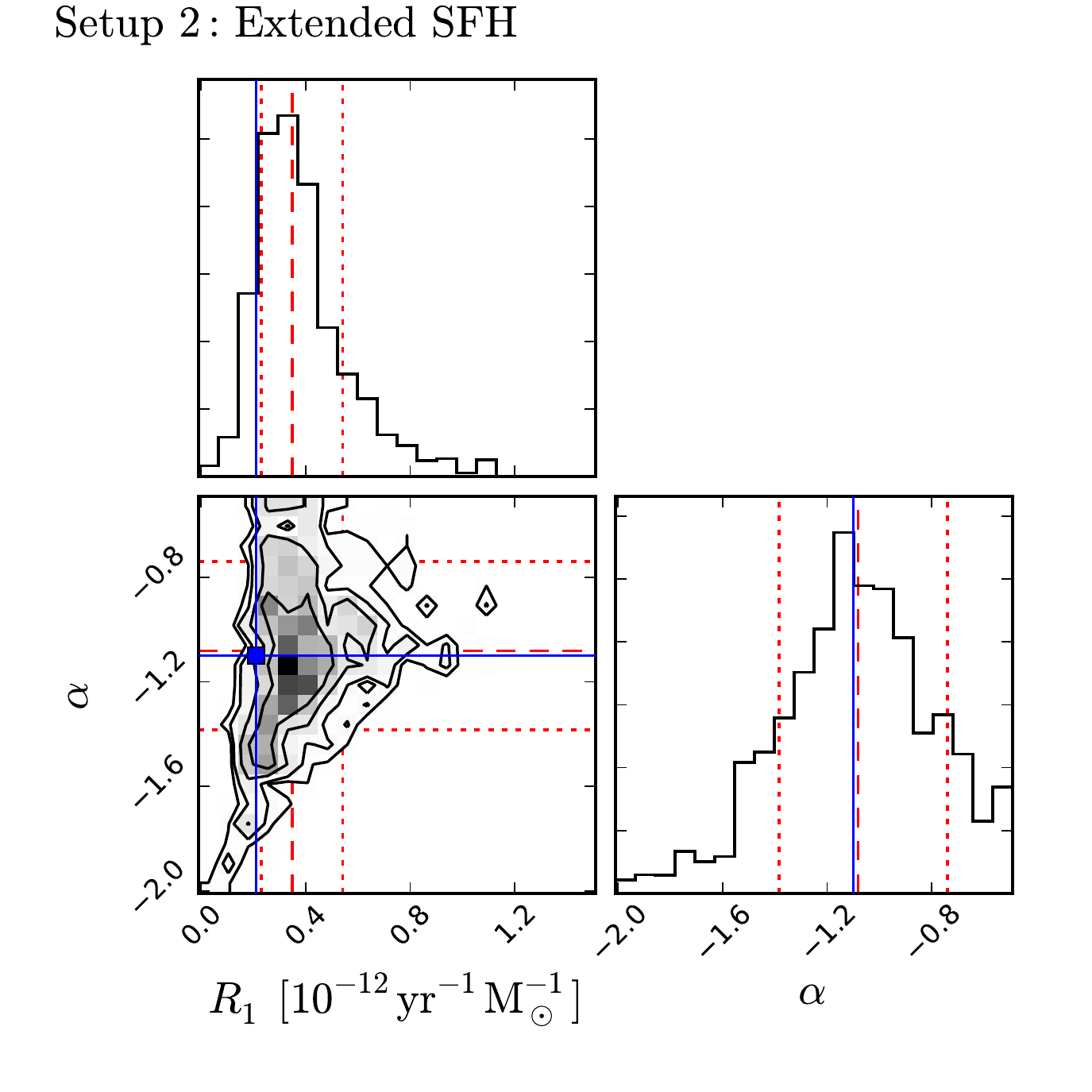}
	\\
	\includegraphics[height=0.4\textwidth,trim={0.2cm 0.5cm 0.cm -0.3cm},clip]{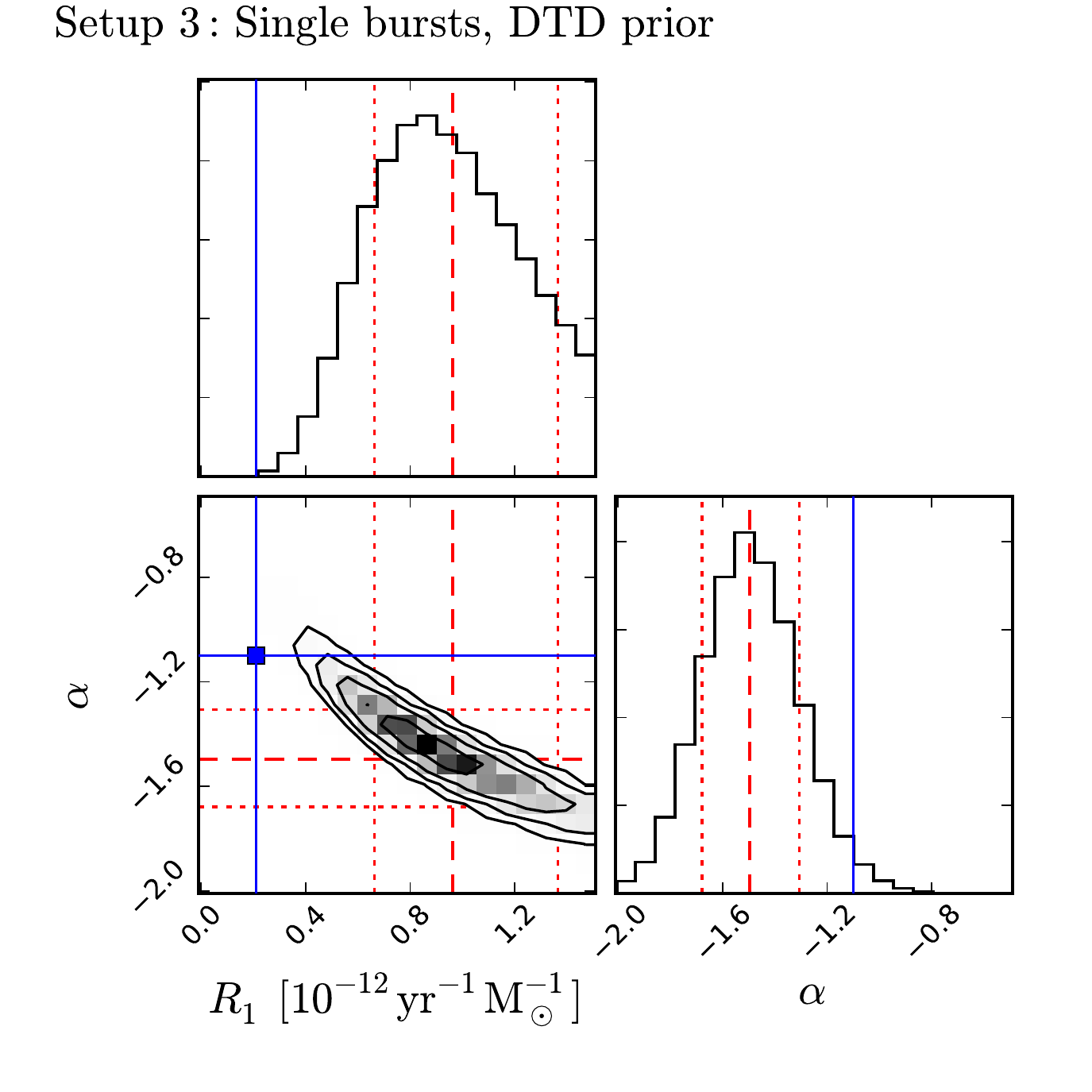}
	\hspace{0.6cm}
	\frame{\includegraphics[height=0.4\textwidth,trim={0.2cm 0.5cm -0.2cm -0.3cm},clip]{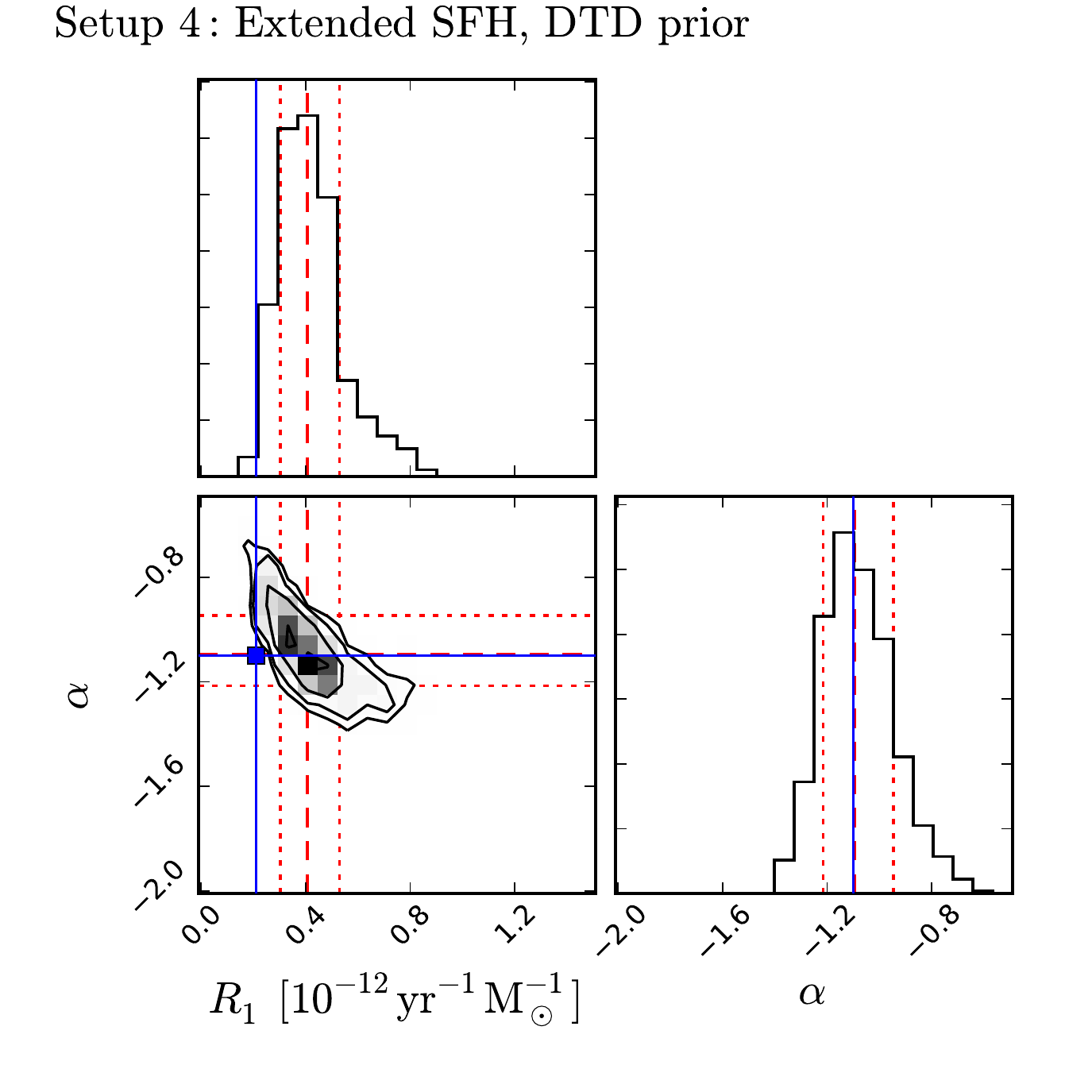}}
	\vspace{-0.1cm}
	\caption{
	Posterior distributions of the DTD parameters $R_1$ and $\alpha$ for the four different setups outlined in section~\ref{subsection:setups}, paralleling Fig.~\ref{fig:corner}. Red dashed and dotted lines indicate the medians, and the $\pm1\sigma$ $16^{\rm th}$ and $84^{\rm th}$ percentiles. The blue square highlighted with a solid cross corresponds to the field-galaxy DTD parameters from \protect\cite{Maoz2017}. Contours correspond to 0.5, 1, 1.5, and 2 $\sigma$ levels. 
	The upper panels illustrate how, when considering only the \FM18 cluster SNe, enabling extended SFH significantly reduces the allowed range for the DTD parameters. The lower panels show that the addition also of priors based low-redshift cluster rate measurements further constrains the DTD parameters. In the lower-right panel, allowing both extended SFH and prior constraints leads to our main result: a DTD power-law slope $\alpha$ consistent with the one measured in field-galaxy environments, and with an amplitude significantly ($3.8\sigma$) higher, by a factor of 2-3, than the field-galaxy SN Ia DTD. 
	}
	\label{fig:corner}
\end{figure*}

Fig.~\ref{fig:DTD} shows the power-law DTDs resulting from the four different MCMC model-fitting setups outlined in section~\ref{subsection:setups}; Fig.~\ref{fig:corner} gives, for each setup, the corresponding posterior distribution of the DTD parameters, $R_1$ and $\alpha$.
The upper panels show the DTDs resulting from fitting only  the photometry and SN Ia numbers of the \FM18 high-redshift cluster sample (setups 1 and 2), while ignoring constraints based on previous cluster SN Ia rate measurements. The lower panels, in contrast, do take into account as a prior the low-$z$ constraints on the DTD parameters (setups 3 and 4). 
The two left panels show the DTDs obtained when assuming single star-formation burst histories for all clusters, as in previous studies; the right panels give the DTDs resulting when allowing, in the \FM18 clusters, extended SFHs, the focus of this work.
For reference purposes alone, in Fig.~\ref{fig:DTD} we plot, beside the model DTDs, the SN Ia rates per unit formed stellar mass (assuming a single starburst) from section \ref{subsection:revised_SNIa}, for the \FM18 sample and for the low-$z$ cluster samples, all as revised in the present paper (identical values are plotted for all of these rates in all four panels). We emphasize, however, that in none of these plots/setups are these sample- and sub-sample SN rates the observables that are being fit by the models. Rather, the observables being fit are the luminosities in the observed bands of each of the \FM18 clusters, and the number of SNe Ia detected within each individual \FM18 cluster.    
In the figures, the MCMC results are also compared to the cluster DTD previously obtained by \FM18 and to the field-galaxy DTD from \citet{Maoz2017}.

The upper left panels in Fig.~\ref{fig:DTD} and in Fig.~\ref{fig:corner} show that, when considering only the \citetalias{Friedmann2018} clusters, with their limited range of short delay times around 2.5 Gyr, the assumption of single star-formation bursts (setup 1) leads to only weak constraints on the DTD parameters, with large covariance between the two values of the two DTD power-law parameters. This result largely echoes the results shown previously in Fig.~\ref{fig:Plow}, for a similar reason---a limited range of short delays (previously long delays). Allowing extended SFHs for the FM18 clusters, however (setup 2, upper right panels in Figs.~\ref{fig:DTD} and ~\ref{fig:corner}), significantly reduces both the allowed range for the DTD parameters and their correlation in the posterior distribution. This outcome already highlights the importance of considering extended SFHs for the high-$z$ clusters with their corresponding short delay times. The SFH parameters of the walkers span most of the allowed ranges in setup 2 (except for the parameter $m$ describing the second star formation burst, for which values above 0.1 are generally excluded).  With an exponential star-formation-rate timescale $\tau$ that can be as high 1.2 Gyr, the clusters are consistent with being observed within less than two star-formation e-folding times, effectively probing a larger range of DTD delay times. Remarkably, the DTD parameters obtained with setup 2, namely 
$R_1=0.35_{-0.12}^{+0.19}\times 10^{-12}~ 
\rm yr^{-1}M_\odot^{-1}$ and 
$\alpha=-1.08_{-0.30}^{+0.34}$, 
are already quite similar to (although more uncertain than) those we obtain from the final setup 4, described below, that includes the leverage provided by the lower-redshift cluster SN rates. This highlights a  consistency between the DTD parameters indicated by the low-$z$ and high-$z$ cluster rates.

In the lower-left panels (setup 3) of Figs.~\ref{fig:DTD} and \ref{fig:corner} one sees that the addition of the low-$z$ SN rate data through the prior $\mathcal{P}_{\rm low}$ constrains better the DTD parameters, shrinking the allowed parameter space. More importantly, the setup 3 calculation is, in essence, a repeat of the analysis of \FM18 (and other previous cluster SN Ia DTD analyses) that assumed single-burst cluster SFHs, only here with our revised high-$z$ SN Ia numbers and cluster luminosities, and the revised low-$z$ cluster SN Ia rates. Unsurprisingly, the best-fit DTDs indeed go through the data points that are plotted for reference, data points that as noted, 
are based on the same single-burst assumption. With these revised numbers and the single-burst assumption, the DTD amplitude remains high as in \FM18 (i.e. several times higher than the field-DTD values), and the DTD power-law index, $\alpha=-1.50_{-0.18}^{+0.19}$, is even steeper than its $ \alpha=-1.30_{-0.16}^{+0.23}$ value in \FM18, which in turn is steeper than the field DTD value. The results of setup 3 thus confirm and strengthen the conclusions of \FM18, of a high-normalised and steep cluster-environment DTD, when assuming single-burst cluster SFHs. 

Finally, in the lower-right panels of Figs.~\ref{fig:DTD} and \ref{fig:corner} (setup 4), we present the main results of this work---the cluster-environment DTD as constrained by the \FM18-sample cluster and SN Ia data, and by the prior lower-$z$ cluster-sample SN Ia rates, while allowing, for the \FM18 clusters,  extended SFHs that are consistent with the data. We obtain
$R_1 = 0.41_{-0.10}
^{+0.12} \times 10^{-12}~ 
\rm yr^{-1}M_\odot^{-1}$ and 
$\alpha=-1.09_{-0.12}^{+0.15}$. In terms of Hubble-time-integrated SN Ia number per formed stellar mass, this corresponds to $N/M_{\rm star} = 3.1_{-1.0}^{+1.1} \times 10^{-3} M_\odot^{-1}$. 
On the one hand, these parameter values are within 1$\sigma$ of the \citetalias{Friedmann2018} values, thus confirming with a different methodology, with revised data and with more realistic SFH assumptions, the \FM18 conclusion that the SN Ia DTD in cluster environments is higher-normalised than in field galaxies. The field-galaxy DTD parameters are $R_1 = 0.21 \pm 0.02 \times 10^{-12}~ \rm yr^{-1}M_\odot^{-1}$, $\alpha=-1.07 \pm 0.09$, and $N/M_{\rm star} = 1.3 \pm 0.1 \times 10
^{-3} M_\odot^{-1}$ \citep{Maoz2017}. On the other hand, the dichotomy in DTD normalisation between the two environments is decreased compared to the results in \FM18, and is now only at the level of a factor of $\sim 2-3$. Furthermore, the DTD power-law index $\alpha$ now appears to be fully consistent, at $\alpha\approx -1.1$, in both cluster and field environments, rather than being steeper in clusters. From comparison of the setup 3 and setup 4 results, it is clear that this convergence in DTD parameters comes about from the more-physical assumption of extended cluster SFHs that we have adopted in this work. Nevertheless the difference between the cluster and field DTD parameters remains significant. From the difference in log likelihood between the 
best fit (setup 4) cluster DTD solution and that for the parameters of the field-galaxy DTD, we deduce a $3.8\sigma$ significance.
 
We summarize in Table~\ref{table:DTD_params} the DTD parameters obtained with MCMC setups 3 and 4, and those from \FM18 and \cite{Maoz2017}. In Table~\ref{table:SFH_params}, we list the SFH parameters for the individual \FM18 clusters from the MCMC fit for the adopted setup 4 calculation.
For each cluster, we list also its model fluxes (which can be compared to Table 1) and the predicted vs. observed SN Ia numbers.

\begin{table}
	\caption{SN Ia DTD parameters.}
	\begin{tabular*}{\linewidth}{l@{\extracolsep{\fill}}c@{\extracolsep{\fill}}c@{\extracolsep{\fill}}c}
		\hline
		\hline
		\noalign{\vskip 0.5mm} 
        Setup & $\alpha$ & $R_1$ $\rm [10^{-12}~yr^{-1}M_\odot^{-1}]$ & $N/M_{\rm star}$ $\rm [10^{-3}~M_\odot^{-1}]$\\
        (1) & (2) & (3) & (4)\\
		\noalign{\vskip 0.5mm} 
		\hline
		\noalign{\vskip 0.5mm} 
		\multicolumn{4}{c}{Cluster-environment DTD (this work)}\\
		\noalign{\vskip 0.5mm} 
		\hline
		\noalign{\vskip 0.8mm} 
	 	 Setup 3 & $-1.50_{-0.18}^{+0.19}$ & $0.96_{-0.30}^{+0.40}$ & $11.3_{-5.0}^{+6.8}$\\
		\noalign{\vskip 0.8mm} 
		{\bf Setup 4 }& $\mathbf{-1.09_{-0.12}^{+0.15}}$& $\mathbf{0.41_{-0.10}^{+0.12}}$ &  $\mathbf{3.1_{-1.0}^{+1.1}}$ \\
		\noalign{\vskip 0.8mm} 
		\hline
		\noalign{\vskip 0.5mm} 
		\multicolumn{4}{c}{Cluster-environment DTD (\citetalias{Friedmann2018})}\\
		\noalign{\vskip 0.5mm} 
		\hline
		\noalign{\vskip 0.8mm} 
		-- & $-1.30_{-0.16}^{+0.23}$ & $0.7_{-0.2}^{+0.1}$ & $5.4_{-2.3}^{+2.3}$ \\
		\noalign{\vskip 0.8mm} 
		\hline
		\noalign{\vskip 0.5mm} 
		\multicolumn{4}{c}{Field-galaxy DTD \citep{Maoz2017}}\\
		\noalign{\vskip 0.5mm} 
		\hline
		\noalign{\vskip 0.8mm} 
		-- & $-1.07 \pm 0.09$ & $0.21 \pm 0.02$ & $1.3 \pm 0.1$ \\
		\noalign{\vskip 0.8mm} 
		\hline
	\end{tabular*}
	\begin{minipage}{1\linewidth}
	\textbf{Notes.} 
	(1) 
	MCMC calculation setup;
	setup 3 assumes single-burst star-formation histories (SFHs), as in previous cluster DTD studies; setup 4, which leads to our final result, allows for more realistic, extended, SFHs in the higher-redshift \FM18 clusters. 
	(2) DTD power-law index $\alpha$, as expressed in Eq.~(\ref{eq:DTD}). 
	(3) DTD normalisation $R_1$, which is the SN Ia rate  per unit formed stellar mass at 1~Gyr delay time. 
	(4) Hubble-time-integrated SN~Ia number per formed stellar mass. 
    \end{minipage}
	\label{table:DTD_params}
\end{table}

\begin{table*}
	\caption{SFH parameters and SNe Ia numbers resulting from setup 4 (with extended SFHs and a DTD prior).}
	\begin{tabular*}{\linewidth}{l@{\extracolsep{\fill}}c@{\extracolsep{\fill}}c@{\extracolsep{\fill}}c@{\extracolsep{\fill}}c@{\extracolsep{\fill}}c@{\extracolsep{\fill}}c@{\extracolsep{\fill}}c@{\extracolsep{\fill}}c@{\extracolsep{\fill}}c@{\extracolsep{\fill}}c}
		\hline
		\hline
		\noalign{\vskip 0.5mm} 
        Cluster name & $t$ & $\tau$ & $A_V$ & $\log(m)$ & $M_0$ & $f_{\rm F105W, model}$ & $f_{\rm F140W, model}$ & $f_{\rm F160W, model}$ & $N_{\rm Ia, model}$ & $N_{\rm Ia, obs}$\\
        (1) & (2) & (3) & (4) & (5) & (6) & (7) & (8) & (9) & (10) & (11) \\
		\noalign{\vskip 0.8mm} 
		\hline
		\noalign{\vskip 0.8mm} 
		IDCS1426   & $1.76_{-0.19}^{+0.20}$ & $0.6_{-0.4}^{+0.4}$ & $0.73_{-0.45}^{+0.45}$ & $-2.36_{-1.14}^{+0.90}$ & $0.81_{-0.35}^{+0.35}$ & $0.72_{-0.24}^{+0.24}$ & $0.63_{-0.02}^{+0.02}$ & $0.57_{-0.05}^{+0.05}$ & $1.6_{-0.6}^{+0.9}$ & $1.5$ \\
		\noalign{\vskip 0.8mm} 
        ISCS1432   & $2.60_{-0.20}^{+0.19}$ & $0.5_{-0.3}^{+0.4}$ & $0.57_{-0.36}^{+0.46}$ & $-2.56_{-0.57}^{+0.37}$ & $1.22_{-0.29}^{+0.29}$ & $0.88_{-0.10}^{+0.10}$ & $0.92_{-0.01}^{+0.01}$ & $0.90_{-0.04}^{+0.04}$ & $1.7_{-0.5}^{+0.9}$ & $1.0$ \\
        \noalign{\vskip 0.8mm} 
        MOO1014    & $2.95_{-0.19}^{+0.19}$ & $0.7_{-0.4}^{+0.4}$ & $0.58_{-0.41}^{+0.55}$ & $-1.91_{-0.32}^{+0.23}$ & $1.37_{-0.32}^{+0.32}$ & $2.16_{-0.15}^{+0.15}$ & $1.67_{-0.08}^{+0.08}$ & $1.51_{-0.11}^{+0.11}$ & $2.3_{-0.7}^{+1.1}$ & $2.0$ \\
        \noalign{\vskip 0.8mm} 
        MOO1142    & $3.32_{-0.23}^{+0.18}$ & $0.6_{-0.4}^{+0.3}$ & $0.60_{-0.40}^{+0.55}$ & $-2.20_{-0.33}^{+0.08}$ & $1.99_{-0.46}^{+0.46}$ & $2.31_{-0.00}^{+0.00}$ & $2.11_{-0.01}^{+0.01}$ & $2.01_{-0.05}^{+0.05}$ & $1.2_{-0.4}^{+0.5}$ & $0.0$ \\
        \noalign{\vskip 0.8mm} 
        SPARCS0224 & $2.11_{-0.21}^{+0.20}$ & $0.7_{-0.4}^{+0.4}$ & $0.65_{-0.37}^{+0.51}$ & $-2.07_{-0.90}^{+0.46}$ & $1.07_{-0.43}^{+0.43}$ & $0.98_{-0.16}^{+0.16}$ & $0.85_{-0.07}^{+0.07}$ & $0.78_{-0.11}^{+0.11}$ & $0.8_{-0.3}^{+0.5}$ & $1.5$ \\
        \noalign{\vskip 0.8mm} 
        SPARCS0330 & $2.12_{-0.19}^{+0.18}$ & $0.7_{-0.3}^{+0.3}$ & $0.58_{-0.43}^{+0.50}$ & $-1.82_{-0.76}^{+0.52}$ & $0.74_{-0.32}^{+0.32}$ & $0.98_{-0.16}^{+0.16}$ & $0.75_{-0.09}^{+0.09}$ & $0.66_{-0.13}^{+0.13}$ & $1.0_{-0.4}^{+0.7}$ & $0.5$ \\
        \noalign{\vskip 0.8mm} 
        SPARCS1049 & $1.88_{-0.19}^{+0.20}$ & $0.8_{-0.4}^{+0.3}$ & $0.77_{-0.56}^{+0.46}$ & $-1.67_{-0.40}^{+0.15}$ & $1.01_{-0.35}^{+0.35}$ & $1.30_{-0.12}^{+0.12}$ & $0.93_{-0.02}^{+0.02}$ & $0.80_{-0.01}^{+0.01}$ & $0.5_{-0.2}^{+0.4}$ & $0.0$ \\
        \noalign{\vskip 0.8mm} 
        SPARCS0035 & $2.81_{-0.18}^{+0.19}$ & $0.7_{-0.4}^{+0.3}$ & $0.61_{-0.36}^{+0.42}$ & $-2.32_{-0.87}^{+0.17}$ & $1.34_{-0.30}^{+0.30}$ & $1.28_{-0.00}^{+0.00}$ & $1.22_{-0.03}^{+0.03}$ & $1.16_{-0.03}^{+0.03}$ & $1.4_{-0.4}^{+0.7}$ & $0.5$ \\
        \noalign{\vskip 0.8mm} 
        SPT0205    & $2.84_{-0.16}^{+0.21}$ & $0.5_{-0.3}^{+0.4}$ & $0.75_{-0.43}^{+0.43}$ & $-2.73_{-0.63}^{+0.47}$ & $1.61_{-0.40}^{+0.40}$ & $1.00_{-0.12}^{+0.12}$ & $1.11_{-0.01}^{+0.01}$ & $1.09_{-0.04}^{+0.04}$ & $2.9_{-0.9}^{+1.4}$ & $4.5$ \\
        \noalign{\vskip 0.8mm} 
        SPT2040    & $2.39_{-0.19}^{+0.17}$ & $0.7_{-0.4}^{+0.3}$ & $0.69_{-0.41}^{+0.43}$ & $-2.45_{-0.68}^{+0.42}$ & $2.31_{-0.77}^{+0.77}$ & $1.76_{-0.27}^{+0.27}$ & $1.69_{-0.08}^{+0.08}$ & $1.62_{-0.15}^{+0.15}$ & $3.3_{-1.0}^{+1.5}$ & $3.0$ \\
        \noalign{\vskip 0.8mm} 
        SPT2106    & $3.46_{-0.17}^{+0.14}$ & $0.7_{-0.4}^{+0.4}$ & $0.67_{-0.50}^{+0.52}$ & $-1.32_{-0.15}^{+0.12}$ & $1.04_{-0.32}^{+0.32}$ & $3.55_{-0.04}^{+0.04}$ & $1.98_{-0.00}^{+0.00}$ & $1.67_{-0.03}^{+0.03}$ & $1.9_{-0.6}^{+1.1}$ & $1.0$ \\
        \noalign{\vskip 0.8mm} 
        XMM44      & $2.17_{-0.17}^{+0.21}$ & $0.7_{-0.4}^{+0.3}$ & $0.72_{-0.44}^{+0.43}$ & $-2.37_{-0.62}^{+0.29}$ & $1.81_{-0.50}^{+0.50}$ & $1.31_{-0.07}^{+0.07}$ & $1.28_{-0.03}^{+0.03}$ & $1.22_{-0.04}^{+0.04}$ & $1.3_{-0.5}^{+0.6}$ & $1.0$ \\
        \noalign{\vskip 0.8mm} 
        \hline
	\end{tabular*}
	\begin{minipage}{1\linewidth}
	\textbf{Notes.} 
	(1) Cluster ID; 
	(2) Delay time, in Gyr; 
	(3) Exponential timescale, in Gyr;
	(4) Visible extinction, in mag; 
	(5) Relative mass $m$ of the second burst, with respect to the first burst; 
	(6) Best-fit asymptotically formed stellar mass, in $10^{13}M_\odot$; 
	(7)-(9) Total modeled galaxy cluster flux in the F105W, F140W, and F160W bands, in $\rm 10^{-16} erg~cm^{-2}~s^{-1}~\AA^{-1}$; 
	(10) Modeled SN Ia count; 
	(11) Observed SN Ia count (cf. Eq.~(\ref{eq:Nobs})). 
    \end{minipage}
	\label{table:SFH_params}
\end{table*}

\section{Conclusions}
\label{section:conclusion}

We have re-analysed the data resulting from the near-infrared  HST imaging observations, described and analysed in \FM18, of a sample of galaxy clusters at redshifts $z=1.13-1.75$ that were monitored for SNe. To re-cap, \FM18 discovered the transient events in these data, selected those events consistent with being SNe Ia hosted by the clusters, and performed multi-band galaxy-light photometry for every cluster in the sample. The photometry, combined with single-burst stellar population synthesis models, was used by \FM18 to estimate the initially formed stellar mass of each cluster. Combining the detected SN Ia numbers, the formed stellar masses, the monitoring period lengths, and the detection efficiencies of the experiment, \FM18 derived the SN Ia rate per unit formed mass, for this sample. Finally, \FM18 fit power-law SN Ia DTDs to the new SN Ia rates, combined with the SN Ia rates measured for previous, lower-redshift, cluster samples. Confirming previous results \citep{Maoz2010,Maoz2017}, \FM18 obtained a cluster-environment DTD that is slightly steeper ($\alpha\approx -1.3$) than seen in field-galaxy environments ($\alpha\approx -1.1$) but, more important, considerably higher-normalised (i.e., with a larger amplitude $R_1$) than the field-galaxy DTD. The high normalisation, if real, would  indicate a high time-integrated efficiency of SN Ia production in galaxy clusters. While this result was intriguing, concern remained that it was affected by the single-starburst assumption that it involved. For the high-$z$ \FM18 clusters, observed only a few rest-frame Gyr after initial star formation at $z=3-4$, some ongoing star formation in some of the cluster galaxies is possible and, in fact, evident to some degree directly in the observations (e.g. blue galaxy colors, evidence of dust). Not accounting properly for such star formation may have led, in the \FM18 analysis, to overestimation of the stellar masses, an overestimate of the delay time probed at a given redshift, or both. This, in turn, could potentially influence the derived DTD.

In this paper, we have re-visited the problem, but now relaxing the single-burst assumption for the \FM18 clusters, and instead allowing, for each cluster, extended SFHs that are consistent with its observed HST photometry. In the said photometry, we realized that \FM18 had sometimes inadvertently over-subtracted the light of galaxies that are foreground to the clusters. We therefore re-measured all of the photometry from the HST images, and have reported the revised values, which are somewhat higher than in \FM18. Furthermore, considering the results of follow-up spectroscopy of some of the SN host galaxies, as reported  by \cite{Williams2020}, we have revised (slightly upwards) the numbers of SNe Ia hosted by a few of the clusters. Our revised SN Ia rates per unit formed stellar mass (still under the single-burst assumption), based on the revised photometry and SN Ia numbers, are only mildly different from the rates in \FM18. Finally, we have also reviewed the SN Ia rates per unit mass in previous, lower-$z$ cluster samples analysed by \FM18. Consistently using single-burst SPS models and the mass-to-light ratios that they predict, we have revised the low-$z$ cluster SN Ia rates (typically downward by $\sim 30\%$).  

To constrain the cluster DTD while allowing for extended SFHs, we have fit, using an MCMC process, the observed photometry to the SEDs of simplified but broad families of parametrised SFHs, as  predicted by SPS models (including also dust extinction). Simultaneously, parametrised power-law DTDs were fit to the data, by comparing the number of SNe Ia predicted in a cluster, given its model-assumed SFH, to the actual number observed in that cluster.
In contrast to \FM18, where the total number of SNe Ia in the cluster sample was considered, and the sample's total stellar mass, here we avoid this loss of information due to averaging, by comparing in each cluster, with its SFH and mass constrained by the photometry, the predicted and the observed SN Ia numbers. The treatment using the individual clusters, which span a range of redshifts and hence of delay times, provides further leverage on the DTD. In the MCMC calculation, to account for the measured SN Ia rates in lower-$z$ cluster samples, we have used them as a prior on the DTD parameters. The SN Ia rate data that are being fit have been slightly revised, as described above, and so in order to understand the cause behind any change in the results compared to those of \FM18, we have performed the MCMC fits using several different setups. In particular, 
in one of the setups (No. 3) we have fit for the DTD with the revised data and our new methodology (i.e. fitting the SN Ia numbers in individual clusters, using the low-$z$ cluster SN Ia rates as a prior), but limiting the SFHs of all clusters to a single brief starburst, as in \FM18 and previous studies. In MCMC setup No. 4, which yields our new results, we fit the same data, but now allowing for extended SFHs in the \FM18 clusters, as described above.

The DTD emerging from our setup 3 is similar in amplitude to the DTD of \FM18 (compare, e.g., their height at an intermediate delay of $\sim 5$~Gyr, in Fig.~3, bottom-left panel). Our new DTD for this setup is, however, slightly steeper than the one in \FM18 ($\alpha=-1.50\pm 0.19$, vs. $\alpha=-1.30^{+0.23}_{-0.16}$, respectively). This is presumably a result of the decrease in 
our revised low-$z$ SN Ia rates, which "pull" the DTD down at long delays. This shows that the revised data numbers we have used here do not lead to a DTD notably different from that of \FM18; it is just as high-normalised, compared to the field-galaxy DTD, and if anything, has even a steeper slope, $\alpha$, compared to the field DTD found by \FM18.    

Proceeding to MCMC setup 4 (Fig. 3, bottom-right panel) fitting all of the data while allowing extended SFHs in the \FM18 clusters, we have found that the DTD slope becomes shallower, $\alpha=-1.09_{-0.12}^{+0.15}$, and is now indistinguishable from the slope of the field DTD, $\alpha\approx -1.1$.
The amplitude of the cluster DTD has become lower and better constrained than in \FM18, $R_1 = 0.41_{-0.10}^{+0.12} \times 10^{-12}~ \rm yr^{-1}M_\odot^{-1}$ (which translates to a Hubble-time-integrated SN Ia number per formed stellar mass, $N/M_{\rm star} = 3.1_{-1.0}^{+1.1} \times 10^{-3} M_\odot^{-1}$), compared to  $R_1=0.96_{-0.30}^{+0.40}\times 10^{-12}~ \rm yr^{-1}M_\odot^{-1}$ in the single-burst setup 3. This difference in the resulting DTD shows that, indeed, accounting for extended SFHs in high-z clusters is important for reliably recovering the DTD in cluster environments: the DTD becomes lower than under the single-burst assumption, and flatter---as flat as the field DTD. However, our new  cluster DTD is still significantly higher-normalised, by a factor of 2-3, compared to the field DTD [$R_1 = (0.21\pm0.02) \times 10^{-12}~ \rm yr^{-1}M_\odot^{-1}$, or $N/M_{\rm star} = (1.3\pm0.1) \times 10^{-3} M_\odot^{-1}$], even if the dichotomy between cluster and field DTDs has narrowed. 

From a purely random-statistical perspective, the difference that we have now confirmed, between the cluster- and field-DTD normalisations, is 
significant, and therefore begs physical explanations (see Section 1 for some possibilities). However, we should also remember that the measurement of SN rates is fraught with systematic uncertainties. We have made an effort to estimate those systematics that we know of, and to include them 
in our error budgets, e.g: the uncertainty in SN Ia numbers because of uncertainty in transient classification and in host-galaxy cluster membership; or the uncertainty in cluster luminosity because of the uncertainty in the choice of the image regions that are representative for the subtraction of foreground and background light contributions. Nonetheless, those estimates of systematic uncertainty could be wrong. Additional systematics that we are unaware of could also be in play, perhaps related to the fact that all of the measurements in the process, from SN detection to the characterisation of stellar populations in galaxies, 
pertain to different types of galaxies---massive early-types in clusters, vs. late-types in the field. 
Considering all of the above, we believe that the current evidence for a cluster DTD that is normalised high by a factor of 2-3, compared to the field-galaxy DTD, remains strong and formally significant, but also that  
caution and further study are warranted.


\section*{Acknowledgements}


We thank M. Friedmann and N. Hallakoun for advice concerning the data re-analysis, and 
B. Rocca-Volmerange, M. Fioc, and L. Correia for their guidance with \texttt{P\'egase.3}. 
This work was supported by grants from the Israel Science Foundation (DM), the German Israeli Science Foundation (DM), and the European Research Council (ERC) under the European Union's FP7 Programme, Grant No. 833031 (DM).


\section*{Data Availability}

The HST imaging data used in this work are available through the data archive at the Space Telescope Science Institute. The processed data from this article are available in the article and its tables. 
The codes used for the MCMC modeling are available at \url{https://github.com/JonathanFreundlich/DTD_MCMC}.



\bibliographystyle{mnras}
\bibliography{FM2020_SNIa_DTD} 





\bsp	
\label{lastpage}
\end{document}